%% file: TASEP-LK_main.tex
\documentclass[12pt]{iopart}

\usepackage{iopams}

\usepackage{graphicx}

\newtheorem{lem}{Lemma}
\newtheorem{teo}{Theorem}
\newtheorem{cor}{Corollary}
\newtheorem{cri}{Criterion}

\newcommand{\oneminus}[1]{{#1}^{\prime}}
\newcommand{\dfull}{{p}}
\newcommand{\dbulk}{{q}}
\newcommand{\ddetr}{{r}}
\newcommand{\ddlim}{{s}}

\newcommand{\ddens}{{y}}

\newcommand{\x}{{x}}

\newcommand{\eqref}{\eref}

\begin{document}
\title[Dynamical transition in the TASEP]{Dynamical transition in the TASEP with Langmuir kinetics: mean-field theory}
\author{D Botto$^{1,2}$, A Pelizzola$^{1,2}$,
M Pretti$^{1,3}$ and M Zamparo$^1$}
\address{$^1$ Dipartimento Scienza Applicata e Tecnologia, Politecnico di Torino, Corso Duca degli Abruzzi 24, 10129 Torino, Italy}
\address{$^2$ INFN, Sezione di Torino, via Pietro Giuria 1, 10125 Torino, Italy}
\address{$^3$ Consiglio Nazionale delle Ricerche - Istituto dei Sistemi Complessi (CNR-ISC), Via dei Taurini 19, 00185 Roma, Italy}
\eads{\mailto{davide.botto@polito.it}, \mailto{alessandro.pelizzola@polito.it}, \mailto{marco.pretti@polito.it}, \mailto{marco.zamparo@polito.it}}
\begin{abstract}
We develop a mean-field theory for the \emph{totally asymmetric simple exclusion process} (TASEP) with open boundaries, in order to investigate the so-called \emph{dynamical transition}. 
The latter phenomenon appears as a singularity in the relaxation rate of the system toward its non-equilibrium steady state. 
In the high-density (low-density) phase, the relaxation rate becomes independent of the injection (extraction) rate, at a certain critical value of the parameter itself, and this transition is not accompanied by any qualitative change in the steady-state behavior. 
We characterize the relaxation rate by providing rigorous bounds, which become tight in the thermodynamic limit. 
These results are generalized to the TASEP with Langmuir kinetics, where particles can also bind to empty sites or unbind from occupied ones, in the symmetric case of equal binding/unbinding rates. 
The theory predicts a dynamical transition to occur in this case as well. 
\end{abstract}
\pacs{02.50.Ga, 05.50.+q, 89.75.-k}
\maketitle

\input{TASEP-LK_intro.tex}

\input{TASEP-LK_model.tex}

\input{TASEP-LK_phase-diagram.tex}

\section{The dynamical transition}
\label{sec:dynamical-transition}

Before turning to more technical details, in this section we present a summary of the main original contributions of our work, in particular the determination of the smallest eigenvalue of the relaxation matrix (that is the slowest relaxation rate) and its singular behavior in the infinite-size limit (that is the dynamical transition), in the framework of the mean-field approximation. 
We shall consider only the HD phase, because our investigation is limited to the symmetric TASEP-LK, which preserves the HD-LD duality, so that everything can be rephrased for the LD phase through the symmetry operation mentioned above. 

Let us denote by ${\lambda_{\min}}^{(N)}$ the slowest relaxation rate for a system of size ${N}$ (in this section we emphasize the size-dependence by attaching a superscript). 
The first result is that the infinite-size limit of this relaxation rate takes the following form
\begin{equation}
  {\lambda_{\min}}^{(\infty)} 
  \equiv \lim_{N \to \infty} {\lambda_{\min}}^{(N)}
  = 1 - \frac{{x}_{*}}{{x}_{\circ}} 
  \, ,
  \label{eq:relaxation_rate}
\end{equation}
where 
\begin{equation}
  \frac{1}{{x}_{\circ}} 
  \equiv 2 \sqrt{ (\beta + \Omega) \oneminus{(\beta + \Omega)} }
  \label{eq:definizione_xo}
\end{equation}
and ${x}_{*}$ is determined by the behavior of a real function $f(x;\alpha, \beta+\Omega)$ of the real variable ${x \geq 1}$ and of the model parameters $\alpha$, $\beta$ and $\Omega$. 
The function is defined as follows
\begin{equation}
  f(x) \equiv \sum_{n=1}^{\infty} 
  \left( \sigma_{n+1} - \sigma_{n-1} \right) 
  {v}_{n}(x) {\zeta(x)}^{n}
  \, , 
  \label{eq:definizione_fx_come_serie}
\end{equation}
where
\begin{equation}
  \zeta(x) \equiv x - \sqrt{x^2-1}
  \, . 
  \label{eq:definizione_z}
\end{equation}
The dependence on the model parameters, which we have no longer denoted explicitly, lies in $\sigma_{n}$ and ${v}_{n}(x)$, defined by recursion respectively as
\begin{equation}
  \sigma_{0} \equiv 2 {x}_{\circ} \alpha
  \, , \qquad 
  \sigma_{n+1} \equiv 2 {x}_{\circ} - 1/\sigma_{n}
  \qquad n = 0,1,2,\dots
  \, , 
  \label{eq:definizione_sigma_ricorsiva}
\end{equation}
with ${x}_{\circ}$ defined by \eqref{eq:definizione_xo}, and
\begin{eqnarray}
  {v}_{0}(x) \equiv 0 \, , \qquad {v}_{1}(x) \equiv 1 
  \, , \label{eq:definizione_v_cond_iniz} \\
  {v}_{n+1}(x) \equiv \left( 2x - \sigma_{n+1} + \sigma_{n-1} \right) 
  {v}_{n}(x) - {v}_{n-1}(x)
  \qquad n = 1,2,\dots
  \, . \label{eq:definizione_v}
\end{eqnarray}
The series in \eqref{eq:definizione_fx_come_serie} turns out to converge very quickly, thus being amenable to numerical evaluation with extremely high precision, at a negligible computational cost.
Let us note that the dependence of $f$ on $\beta$ and $\Omega$ goes only through their sum, which points out that, as announced in the previous section, the slowest relaxation rate of the TASEP-LK (HD phase) can be deduced from that of the pure TASEP by the mapping ${\beta \mapsto \beta+\Omega}$. 
It turns out that, given a particular value of ${\beta+\Omega}$, there exists an interval of $\alpha$ values, larger than a critical threshold $\alpha_\mathrm{c}$,\footnote{Of course this $\alpha_\mathrm{c}$ does not coincide with the exact one, mentioned in the previous section.} such that 
\begin{equation}
  f(x;\alpha,\beta+\Omega) < 1
  \qquad \forall x \geq 1
  \label{eq:f_disuguaglianza}
  \, .
\end{equation}
As a consequence of Theorem~\ref{teo:lambda} and Criterion~\ref{cri:uno} (both stated in the next section), in all this interval the relaxation rate is determined by \eqref{eq:relaxation_rate} with ${x_* = 1}$ and thence it is independent of $\alpha$. 
Otherwise, when $\alpha$ becomes smaller than $\alpha_\mathrm{c}$, condition \eqref{eq:f_disuguaglianza} no longer holds, and in particular ${f(1;\alpha,\beta+\Omega) > 1}$. 
The relaxation rate corresponds to \eqref{eq:relaxation_rate} with ${x_* > 1}$, the precise value of $x_*$ being determined by the equation 
\begin{equation}
  f(x_*;\alpha,\beta+\Omega) = 1
  \label{eq:f_equazione}
  \, .
\end{equation}
This is explained by Theorem~\ref{teo:lambda} together with Criterion~\ref{cri:due} (see next section). 
In this case, still at fixed ${\beta+\Omega}$, the $x_*$ value depends on $\alpha$, in particular $x_*$ increases (and thence the relaxation rate decreases) upon decreasing $\alpha$. 
No discontinuity is observed in $x_*$ as a function of $\alpha$. 
The critical threshold $\alpha_\mathrm{c}$ is determined by the equation
\begin{equation}
  f(1;\alpha_\mathrm{c},\beta+\Omega) = 1
  \label{eq:f_equazione_soglia}
  \, .
\end{equation}

\begin{figure}
  \centering
  \resizebox{100mm}{!}{\includegraphics*{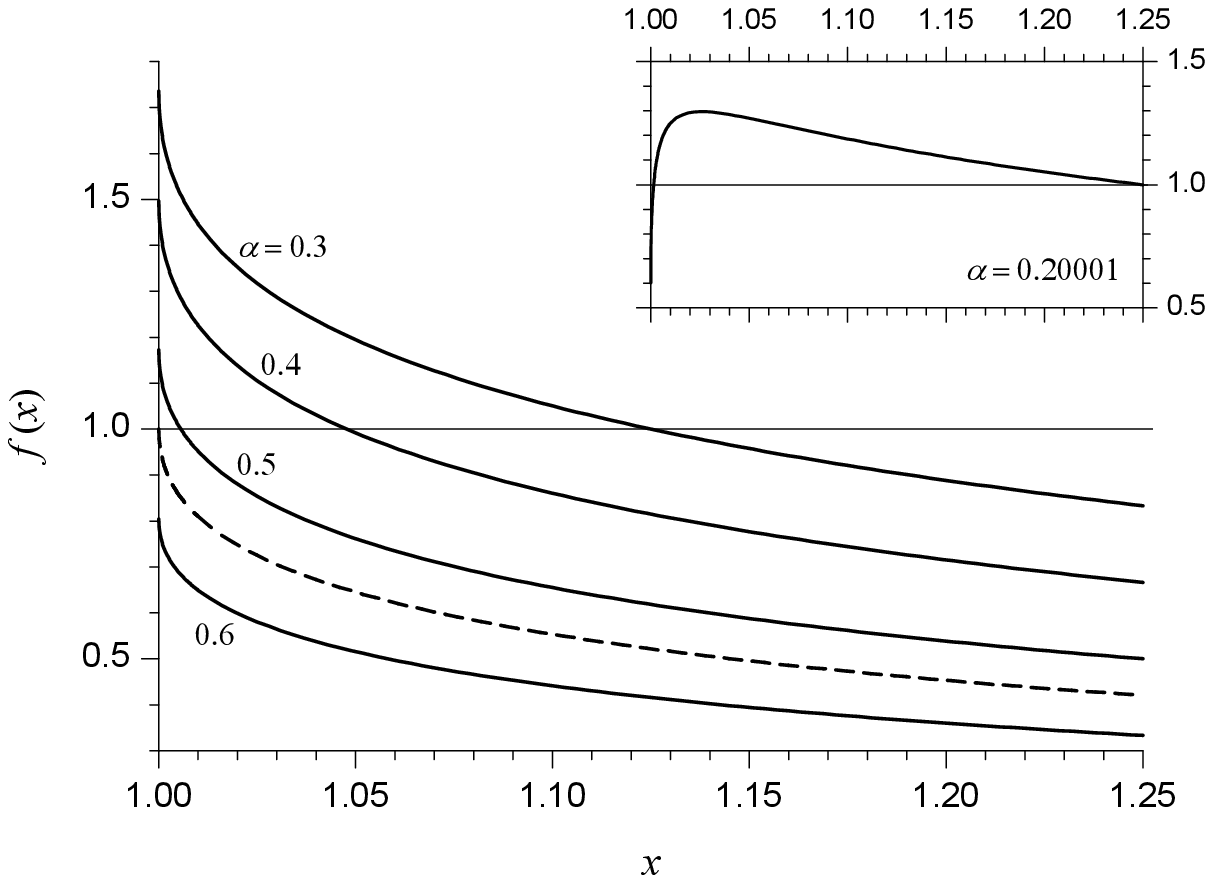}}
  \caption
  {
    Plots of $f(x;\alpha,\beta+\Omega)$ as a function of the variable $x$ in the interval ${[1,x_\circ]}$, for ${\beta+\Omega = 0.2}$ (thence ${x_\circ = 1.25}$) and different $\alpha$ values. 
    The case ${\alpha = \alpha_\mathrm{c}}$ is denoted by a dashed line. 
    The inset displays a very low $\alpha$ case (still in the HD phase). 
  }
  \label{fig:funzione_f}
\end{figure}

All these observations are summarized in figure~\ref{fig:funzione_f}, which displays the behavior of $f$ as a function of $x$ for some representative values of the parameters. 
Note that \eqref{eq:f_equazione_soglia}, at fixed $\Omega$, defines a critical line in the $\alpha$-$\beta$ phase diagram. 
This line and its symmetric counterpart in the LD phase have been previously reported in figures \ref{fig:phase_diagram_TASEP} and \ref{fig:phase_diagram_TASEP-LK}.
Let us also note that in figure \ref{fig:funzione_f} we only consider the interval ${1 \leq x \leq x_\circ}$, because it is possible to prove (see Lemma~\ref{lem:successione_v}, statement \eqref{statement2}) that $x_*$ is bounded therein. 
As previously discussed, at fixed ${\beta + \Omega}$ the lower-bound value ${x_* = 1}$ (corresponding to ${{\lambda_{\min}}^{(\infty)} = 1-1/x_\circ}$) occurs in the whole region ${\alpha \geq \alpha_\mathrm{c}}$. 
On the other hand we can observe (see the inset in figure~\ref{fig:funzione_f}) that $x_*$ tends to the upper-bound value $x_\circ$ (which corresponds to ${{\lambda_{\min}}^{(\infty)} \to 0}$) as $\alpha$ tends (from above) to ${\beta+\Omega}$, that is to the boundary between the pure HD phase and the LD/HD coexistence region. 
At this boundary we indeed expect the onset of an ``ordinary'' dynamical transition (i.e. one accompanied by a static transition) and of a relaxation process being no longer exponential. 
From the inset of figure~\ref{fig:funzione_f} we can also observe that, close to the boundary, the function $f(x)$ is no longer monotonic and a second solution of equation \eqref{eq:f_equazione} may appear. 
This solution can be shown to be irrelevant on the basis of a rigorous argument, still descending from Criterion~\ref{cri:due} (next section).

Let us briefly mention the fact that it is possible to give a physical interpretation of the argument of the square root appearing in the definition of $x_\circ$ \eqref{eq:definizione_xo}, thus generalizing an empirical observation of \cite{PelizzolaPretti17}. 
Let us note that a novelty item of the TASEP-LK with respect to the pure TASEP is that the steady-state current is no longer uniform. 
Nevertheless, it is obviously possible to define a \emph{maximum current}, being still a node-independent quantity. 
Now, denoting by ${{J}_{\max}}^{(N)}$ the maximum current for a system of size ${N}$, the infinite-size limit turns out to be 
\begin{equation}
  {{J}_{\max}}^{(\infty)} 
  \equiv \lim_{N \to \infty} {{J}_{\max}}^{(N)}
  = \frac{1}{4{x_\circ}^2}
  \, ,
\end{equation}
which is precisely the aforementioned square-root argument. 
This result is a consequence of \eqref{eq:bound_current-max} in Corollary~\ref{cor:current} (next section) together with \eqref{eq:SteadyStateMF_detrended_right}. 
The physical meaning is that a larger value of the maximum steady-state current entails (other things being equal) a smaller value of the slowest relaxation rate or, in more general words, that a steady state being more out of equilibrium requires a longer time to be established. 

\begin{figure}
	\centering
	\resizebox{100mm}{!}{\includegraphics*{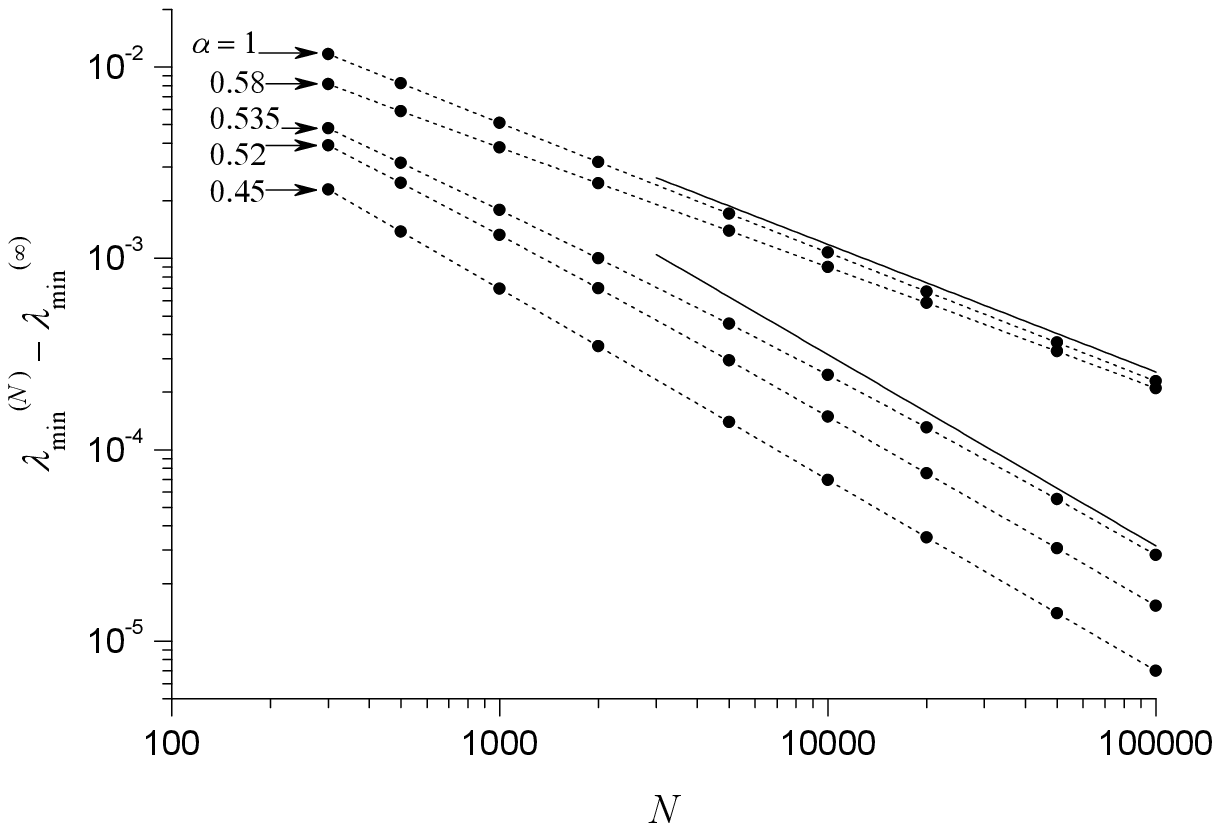}}
	\caption
	{
		Difference between the finite-size slowest relaxation rate ${\lambda_{\min}}^{(N)}$ and its infinite-size limit value ${\lambda_{\min}}^{(\infty)}$ as a function of $N$ (circles) for the TASEP-LK (${\Omega = 0.1}$) with ${\beta = 0.1}$ and different $\alpha$ values, both above and below the critical value ${\alpha_\mathrm{c} \approx 0.54786}$. 
		Solid lines represent scaling functions, i.e. ${\propto N^{-2/3}}$ and ${\propto N^{-1}}$, respectively (see the text).
		Dotted lines are an eyeguide.
	}
	\label{fig:scaling_TASEP-LK}
\end{figure}

As a by-product of the results reported so far, we have also proved bounds for the asymptotic behavior of the relaxation rate ${\lambda_{\min}}^{(N)}$ for ${{N} \to \infty}$. 
In particular it turns out that the dynamical transition discriminates between two different scaling behaviors, which are in turn affected by the presence (or absence) of the Langmuir kinetics, independently of the precise value of the attachment/detachment rate $\Omega$ (provided that nonzero). 
In formulae, for ${\alpha \geq \alpha_\mathrm{c}}$ (${{x}_{*} = 1}$) we have
\begin{equation}
  {\lambda_{\min}}^{(N)} = {\lambda_{\min}}^{(\infty)} + 
  \cases{
	\mathrm{O}\left({N}^{-2}\right)   & if ${\Omega = 0}$ \\
	\mathrm{O}\left({N}^{-2/3}\right) & if ${\Omega > 0}$
  }
  \, ,
\end{equation}
whereas for ${\alpha < \alpha_\mathrm{c}}$ (${{x}_{*} > 1}$) we have
\begin{equation}
  {\lambda_{\min}}^{(N)} = {\lambda_{\min}}^{(\infty)} + 
  \cases{
	\mathrm{O}\left(\zeta({x}_{*})^{2{N}}\right) & if ${\Omega = 0}$ \\
	\mathrm{O}\left({N}^{-1}\right)              & if ${\Omega > 0}$
  }
  \, ,
\end{equation}
where $\zeta({x})$ is defined by \eqref{eq:definizione_z} (satisfying ${\zeta(x) < 1}$ for ${x > 1}$). 
Let us note that the above results, descending from Theorem~\ref{teo:lambda}, are expressed in terms of the ``$\mathrm{O}$'' asymptotic notation, which means that the distance between ${\lambda_{\min}}^{(N)}$ and ${\lambda_{\min}}^{(\infty)}$ is asymptotically bounded \emph{from above} (up to a positive constant factor) by the scaling functions appearing in parentheses. 
This is what one can prove rigorously. 
Indeed, a numerical analysis provides rather clear evidence of the fact that ``$\mathrm{O}$'' \emph{may be replaced by} ``$\Theta$'', meaning that the scaling is optimal. 
\begin{figure}
	\centering
	\resizebox{100mm}{!}{\includegraphics*{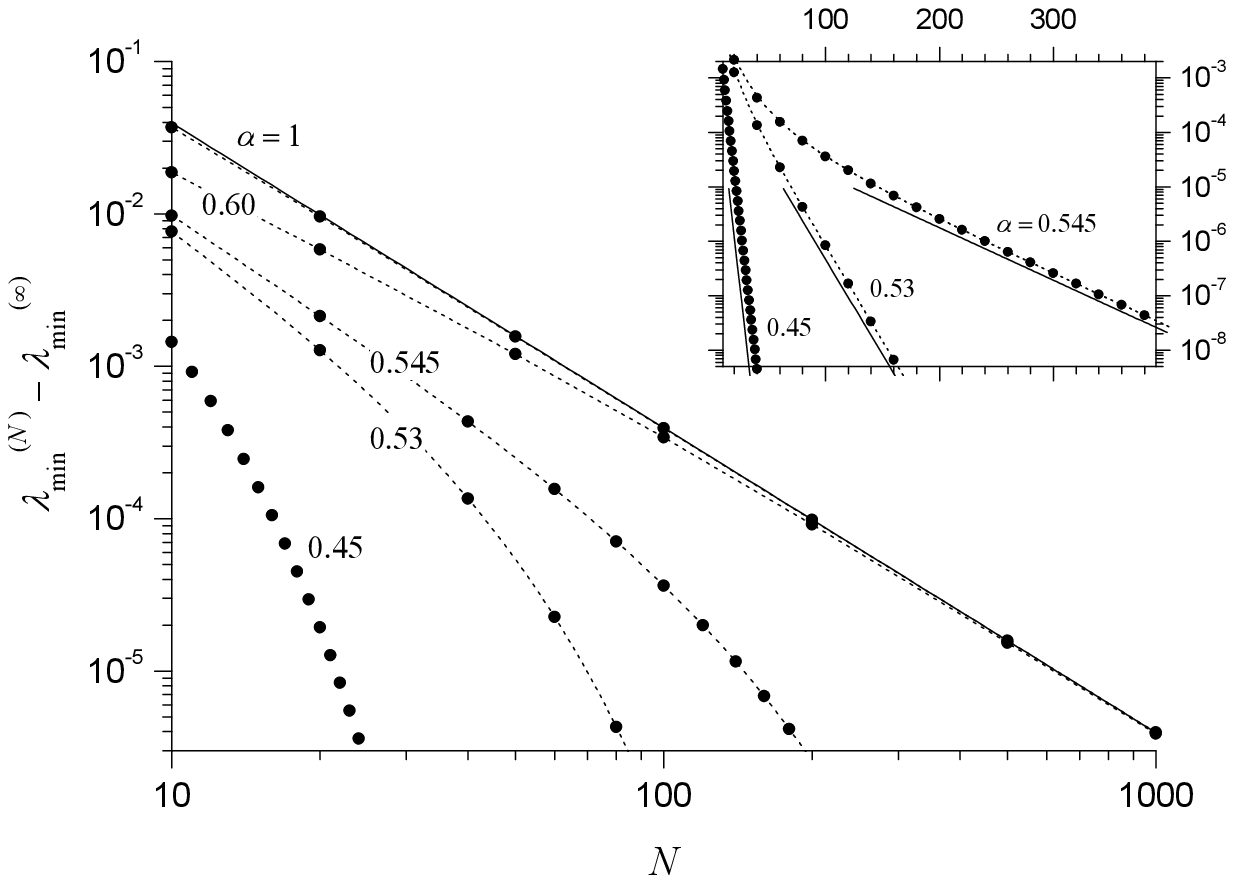}}
	\caption
	{
		Difference between the finite-size slowest relaxation rate ${\lambda_{\min}}^{(N)}$ and its infinite-size limit value ${\lambda_{\min}}^{(\infty)}$ as a function of $N$ (circles) for the pure TASEP (${\Omega = 0}$) with ${\beta = 0.2}$ and different $\alpha$ values, both above and below the critical value ${\alpha_\mathrm{c} \approx 0.54786}$. 
		Solid lines represent scaling functions, i.e. ${\propto N^{-2}}$ in the main figure (with the constant prefactor ${\pi^2 \sqrt{ \beta \oneminus{\beta} }}$, see the text) and ${\propto \zeta(x_*)^{2N}}$ in the inset (with $x_*$ depending on $\alpha$ and $\beta$, see the text).
        Dotted lines are an eyeguide.
	}
	\label{fig:scaling_TASEP}
\end{figure}
A comparison between numerical results and the proposed scaling functions are reported in figures \ref{fig:scaling_TASEP} and~\ref{fig:scaling_TASEP-LK}, respectively for the pure TASEP and the TASEP-LK.
Note that, with our specific choice of parameter values, in both cases the critical injection rate $\alpha_\mathrm{c}$ is the same, because the latter depends only on the sum ${\beta + \Omega}$, as previously discussed. 
As far as the TASEP-LK is concerned (figure~\ref{fig:scaling_TASEP-LK}), we clearly observe that ${{\lambda_{\min}}^{(N)} - {\lambda_{\min}}^{(\infty)}}$ scales with ${N}$ as a power law with two different exponents, depending on whether ${\alpha > \alpha_\mathrm{c}}$ or ${\alpha < \alpha_\mathrm{c}}$. 
For ${\alpha > \alpha_\mathrm{c}}$ (i.e.~in the phase where ${\lambda_{\min}}^{(\infty)}$ does not depend on $\alpha$) the numerical results suggest that the multiplying factor might be independent of $\alpha$ as well.
Concerning the pure TASEP (figure~\ref{fig:scaling_TASEP}), the situation is slightly more heterogeneous, in that for ${\alpha < \alpha_\mathrm{c}}$ the scaling behavior is exponential in $N$, with the characteristic parameter depending on ${x}_{*}$, and thence on $\alpha$. On the other hand, for ${\alpha > \alpha_\mathrm{c}}$ we still have a power law, with a multiplying factor quite clearly independent of $\alpha$, and indeed consistent with the value ${\pi^2 \sqrt{ \beta \oneminus{\beta}}}$, which can be argued from \ref{app:upperbound_mu}. 
Let us note that in principle it is possible that some other scaling form appears for ${\alpha = \alpha_\mathrm{c}}$. 
We have not further investigated about this issue, due to its rather limited interest in the framework of a mean-field theory. 

As far as the maximum current is concerned, \eqref{eq:bound_current-max} in Corollary~\ref{cor:current} along with \eqref{eq:relazione_gamma-zetaxo} point out the following asymptotic behavior 
\begin{equation}
  {{J}_{\max}}^{(N)} = {{J}_{\max}}^{(\infty)} + 
  \cases{
	\mathrm{O}\left(\zeta(x_\circ)^{2N}\right)  & if ${\Omega = 0}$ \\
	\mathrm{O}\left({N}^{-1}\right)    & if ${\Omega > 0}$
  }
  \, ,
\end{equation}
with $x_\circ$ and $\zeta(x)$ still defined by \eqref{eq:definizione_xo} and \eqref{eq:definizione_z}. 
It is worth observing that such a scaling behavior does not undergo any change at the dynamical transition. 
The latter fact confirms the peculiarity of this transition, which does not affect steady-state quantities.

\section{Rigorous results}
\label{sec:rigorous-results}

In this section we present a rigorous derivation of the results reported in the previous one. 
The most technical proofs are further postponed to \ref{app:Rho} and \ref{app:Lambda}. 

\input{TASEP-LK_hd-phase.tex}

\input{TASEP-LK_eigenvalues.tex}

\section{Conclusions and perspectives}
\label{sec:conclusions}

In this paper we have worked out a mean-field theory for the TASEP with open boundaries, possibly with Langmuir kinetics (TASEP-LK), in order to investigate the dynamical transition. 

In the case of pure TASEP, our theory predicts dynamical transition lines being correctly separated by any static transition, and overall in good qualitative agreement with the exact ones~\cite{deGierEssler05,deGierEssler06,deGierEssler08}. 
In particular, in the high-density phase (the low-density phase being equivalent because of the symmetry of the model) the critical value of the injection rate $\alpha_\mathrm{c}$ correctly displays an increasing trend upon decreasing the extraction rate $\beta$, that is upon increasing the bulk density. 
This result was actually unexpected, because a previous mean-field approach~\cite{ProemeBlytheEvans11}, based on the so-called viscous Burgers equation, predicted (agreeing in turn with the domain-wall theory~\cite{KolomeiskySchutzKolomeiskyStraley1998,DudzinskiSchutz2000,NagyAppertSanten2002}) a dynamical transition coinciding with the subphase boundary, namely ${\alpha_\mathrm{c} = 1/2}$ in the high-density phase, and thence completely independent of the extraction rate $\beta$. 
We believe that such a discrepancy is to be ascribed to some extra (uncontrolled) approximation introduced by the continuum limit, from which the description in terms of Burgers equation descends. 
This is indeed the reason why in this paper we have spent a considerable effort to derive all the results in a fully rigorous way, under the initial assumptions of the mean-field approximation, and without resorting to the continuum limit. 

As far as the TASEP-LK is concerned, exact results are not (yet) available, and our analysis, by now limited to the symmetric case of equal binding/unbinding rates, predicts an analogous dynamical transition to take place in the pure high-density and low-density phase regions. 
The latter result suggests that the dynamical transition is likely a robust phenomenon, which is not destroyed by the onset of an additional process, such as the Langmuir kinetics, which establishes a competition with the transport process of the TASEP. 
In this perspective, our theory is a good candidate to serve as a tool for investigating the onset of dynamical transitions in other variants of the TASEP, starting for instance from the TASEP-LK with asymmetric binding/unbinding rates. 
The latter model, for which the cited paper by Parmeggiani and coworkers~\cite{ParmeggianiFranoschFrey04} has pointed out an even richer steady-state phase diagram, might be an interesting subject for an extension of the present theory, which we are indeed meant to develop in the near future.

\appendix

\input{TASEP-LK_appendix-C.tex}

\input{TASEP-LK_appendix-A.tex}

\input{TASEP-LK_appendix-B.tex}

\end{document}

%% file: TASEP-LK_intro.tex
\section{Introduction}
\label{sec:intro}

Non-equilibrium statistical mechanics is an extremely active field in modern theoretical physics. Several decades of intense research efforts have not yet produced a full, well-founded theory like in equilibrium statistical physics, but a lot of results have been obtained so far, either exact or approximate, especially concerning the properties of \emph{non-equilibrium steady states} of various models and phenomena. 

Among many models that exhibit a non-trivial (non-equilibrium) steady state, a paradigmatic one is the \emph{totally asymmetric simple-exclusion process} (TASEP). 
Basically, such a model is made up of a one-dimensional lattice, whose nodes may be occupied by at most one particle, and each particle can hop to the adjacent node, provided the latter is empty. 
The model is called \emph{totally asymmetric}, because hopping is permitted in one direction only. The TASEP was originally introduced in \cite{MacDonaldGibbsPipkin68}, in a more general form, as a model of mRNA translation 
(see \cite{ChouMallickZia11} for a recent review, including several generalizations and applications). 
Various types of boundary conditions have also been considered in the literature. 
In the current paper we shall focus on the specially interesting case of open boundary conditions \cite{Krug91}, where particles are injected at one end of the lattice and extracted at the opposite end, at given rates. 
For this case, a lot of exact results are known, and in particular the steady-state phase diagram in the infinite-size limit, along with a detailed characterization of the various phases and phase transitions. 
We shall consider the pure TASEP as a starting point and, among various possible generalizations, we shall focus on the TASEP with Langmuir kinetics (TASEP-LK), first introduced in \cite{ParmeggianiFranoschFrey03}, where particles can also bind to an empty node or unbind from an occupied node, at given rates. 
In this paper, we limit the analysis to the symmetric case of equal binding/unbinding rates.

Besides the phase transitions affecting the steady state, which we shall denote as \emph{static transitions}, the TASEP is known to exhibit a so-called \emph{dynamical transition}, characterized by a singularity in the relaxation rate. 
The dynamical transition and its precise location in the phase diagram have been studied exactly by de~Gier and Essler \cite{deGierEssler05,deGierEssler06,deGierEssler08} and numerically by Proeme, Blythe and Evans \cite{ProemeBlytheEvans11}. 
A peculiar feature of this transition is that it is not accompanied by any static transition. 
On the other hand, different approximate theories (such as the domain-wall theory~\cite{KolomeiskySchutzKolomeiskyStraley1998,DudzinskiSchutz2000,NagyAppertSanten2002} and the mean-field theory in the continuum limit, i.e.~the viscous Burgers equation~\cite{ProemeBlytheEvans11}) predict a wrong location for the transition, namely coinciding with the high-density and low-density subphase boundaries, which correspond to a qualitative change of the steady-state density profile. 
At odds with the cited results, a recent study by two of us~\cite{PelizzolaPretti17} demonstrates that a mean-field analysis with no further approximations (i.e.~without the continuum limit) turns out to predict a dynamical transition line far from static transitions and closer to the exact one. 
The latter paper also reports a rather clear numerical evidence of a qualitative change in the spectrum of the mean-field relaxation matrix, occurring at the dynamical transition. 

The purpose of the current paper is twofold. 
On the one hand, we shall provide rigorous bounds for the mean-field relaxation rate. 
Such bounds become tight in the infinite-size limit, pointing out a singularity, and thus proving that the mean-field theory actually predicts a dynamical transition at a different location from any static transition. 
A by-product of this result is a very precise characterization of the mean-field steady-state density profile in the high-density phase. The latter is known to be (along with its symmetric low-density counterpart) the one where the dynamical transition takes place. 
Furthermore, we can generalize the results to the aforementioned TASEP-LK model, showing that, at least in the mean-field framework, a dynamical transition takes place in this case as well, and still does not correspond to any static transition. 
Our results help elucidate, to some extent, the physics underlying the dynamical transition, and suggest that the latter is likely to be a rather robust phenomenon, as it is not destroyed by an external process, such as the Langmuir kinetics, which establishes a competition with the hopping process of the TASEP. 
In this perspective, we believe that our analytical approach might also be of use for investigating the onset of a dynamical transition in a variety of models derived from the TASEP, for which an exact solution is not (yet) available.

The paper is organized as follows. In section~\ref{sec:model} we introduce the model and the related mean-field theory, whereas section~\ref{sec:phase-diagram} is a summary of known results about the phase diagram of the TASEP, with and without Langmuir kinetics. 
Section~\ref{sec:dynamical-transition} is the central one, where we describe our results concerning the dynamical transition in the mean-field framework. 
In section~\ref{sec:rigorous-results} the same results are derived in rigorous terms. 
Subsection~\ref{sec:hd-phase} reports in detail the mean-field solution for the density profiles in the high-density phase. 
This subsection is preparatory to \ref{sec:relaxation-rates}, in which we state the rigourous bounds for the relaxation rate, and identify the dynamical transition for both the pure TASEP and the TASEP-LK. 
In section~\ref{sec:conclusions} we draw some conclusions and outline possible future investigations. 
A number of proofs that are omitted in the main text are reported in three appendices.

%% file: TASEP-LK_model.tex
\section{The model and the mean-field theory}
\label{sec:model}

The model is defined on a one-dimensional lattice of $N$ nodes, where particles can hop only rightward, from each node to the adjacent one, provided the latter is empty, at unit rate. 
Particles are injected at the leftmost node (provided it is empty) with rate $\alpha$, and extracted from the rightmost node (provided it is occupied) with rate $\beta$. 
In the TASEP-LK, additional binding/unbinding processes (Langmuir kinetics) are allowed, namely, a particle can bind to an empty node with rate $\omega_\mathrm{A}$ (attachment rate) or unbind from an occupied node with rate $\omega_\mathrm{D}$ (detachment rate). 
As observed in \cite{ParmeggianiFranoschFrey03}, in the limit of large $N$, the physically interesting case is the one in which a competition can be established between the hopping process and the Langmuir kinetics, that is, when the attachment and detachment rates scale as the inverse of $N$. 
For this reason, in the following we shall define $\omega_\mathrm{A} \equiv \Omega_\mathrm{A}/(N+1)$ and $\omega_\mathrm{D} \equiv \Omega_\mathrm{D}/(N+1)$, and assume that the ``macroscopic'' rates $\Omega_\mathrm{A}$ and $\Omega_\mathrm{D}$ are independent of the system size. 
More precisely, since we shall restrict to the symmetric case, we shall impose ${\Omega_\mathrm{A} \equiv \Omega_\mathrm{D} \equiv \Omega}$ and thence ${\omega_\mathrm{A} \equiv \omega_\mathrm{D} \equiv \omega}$. 
The pure TASEP is recovered for ${\Omega = 0}$. 

The steady state of the TASEP with open boundaries was worked out exactly in the 1990s \cite{DerridaDomanyMukamel92,SchutzDomany93,Derrida-etal93,Derrida98}, whereas the mean-field theory was discussed in \cite{DerridaDomanyMukamel92}. 
Conversely, we are not aware of any exact solution for the TASEP-LK, which has been studied (so far only at the steady-state level) by mean-field approximations and numerical simulations \cite{ParmeggianiFranoschFrey03,Popkov03,Evans03,ParmeggianiFranoschFrey04}. 
The mean-field theory for the TASEP-LK (and for the pure TASEP as a special case) can be summarized as follows. 
Let us label lattice nodes from left to right by ${n = 1,\dots,N}$ and let $\dfull_{n}(t)$ denote the \emph{occupation probability} (which we shall also call \emph{local density}) of node $n$ at time $t$. 
Steady-state occupation probabilities, which no longer depend on time, will be simply denoted as $\dfull_{n}$. 
The boundary conditions can be represented by two extra nodes ${n=0}$ and ${n=N+1}$ of fixed ``densities''\footnote{Here we avoid the term \emph{probabilities}, because in principle $\dfull_{0}$ might be larger than $1$ (if ${\alpha > 1}$), whereas $\dfull_{N+1}$ might be less than $0$ (if ${\beta > 1}$).} 
\begin{eqnarray}
  \dfull_{0} & = \alpha 
  \, , 
  \label{eq:LocalDensityLeft}
  \\
  \dfull_{N+1} & = 1-\beta \equiv \oneminus{\beta}
  \, . 
  \label{eq:LocalDensityRight}
\end{eqnarray}
Note that in equation~\eqref{eq:LocalDensityRight} we have introduced the notation ${\oneminus{x} \equiv 1-x}$, which will often allow us to write shorter and clearer formulae in the following.
In general, occupation probabilities have to obey a set of continuity equations, which, taking into account both hopping and Langmuir kinetics, read
\begin{equation}
  \dot{\dfull}_{n}(t)
  = {J}_{n-1}(t) - {J}_{n}(t)
    + \omega_\mathrm{A} \oneminus{\dfull_{n}}(t)
    - \omega_\mathrm{D} \dfull_{n}(t)
  \qquad n = 1,\dots,N
  \, ,
  \label{eq:LocalDensityTimeEvo}
\end{equation}
where ${J}_{n}(t)$ denotes the probability current from node $n$ to ${n+1}$.
Since the hopping rate is set at unity by construction, this current simply coincides with the \emph{joint} probability that node $n$ is occupied \emph{and} node ${n+1}$ is empty. 
The mean-field approximation amounts to treat the two joint events as if they were independent, that is, to impose the current-density relationship
\begin{equation}
  {J}_{n}(t) \equiv \dfull_{n}(t) \, \oneminus{\dfull_{n+1}}(t)
  \qquad n = 0,\dots,N
  \, .
  \label{eq:Current-Density}
\end{equation}
This way, \eqref{eq:LocalDensityTimeEvo} and~\eqref{eq:Current-Density} provide a closed set of equations, which will be the starting point of our subsequent analysis. 
In particular, we shall consider the equations for the steady-state density profile
\begin{equation}
  \dfull_{n} \left( \oneminus{\dfull_{n+1}} + \omega_\mathrm{D} \right)
  = \left( \dfull_{n-1} + \omega_\mathrm{A} \right) \oneminus{\dfull_{n}}
  \qquad n = 1,\dots,N
  \, ,
  \label{eq:SteadyStateMF}
\end{equation}
obtained imposing in \eqref{eq:LocalDensityTimeEvo} the stationarity condition ${\dot{\dfull}_{n}(t) = 0}$. 
The latter equations can be solved numerically, by recasting them in a fixed-point form, namely 
\begin{equation}
  \dfull_{n} = \left( 1 + \frac
    {\oneminus{\dfull_{n+1}} + \omega_\mathrm{D}}
    {\dfull_{n-1}\hphantom{'} + \omega_\mathrm{A}} 
  \right)^{-1}
  \qquad n=1,\dots,N
  \, ,
  \label{eq:SteadyStateMF_iteration}
\end{equation}
while keeping $\dfull_{0}$ and $\dfull_{N+1}$ fixed.
Assuming that the rates $\alpha$ and $\beta$ are both strictly positive, it is possible to prove (see \ref{app:EU}) that \eqref{eq:SteadyStateMF} with the boundary conditions \eqref{eq:LocalDensityLeft} and~\eqref{eq:LocalDensityRight} admits a unique solution with the property 
\begin{equation}
  0 < \dfull_{n} <1 \qquad n=1,\dots,N 
  \, . 
  \label{eq:SteadyStateMF_bound}
\end{equation}
In order to investigate the relaxation process, we now linearize \eqref{eq:LocalDensityTimeEvo}-\eqref{eq:Current-Density} around the steady state. 
In practice, we assume the local densities to be the stationary ones $\dfull_{n}$ plus a small time-varying perturbation $\ddens_{n}(t)$, namely 
\begin{equation}
  \dfull_{n}(t) = \dfull_{n} + \ddens_{n}(t) 
  \qquad n = 0,\dots,N+1
  \, ,
  \label{eq:Perturbation}
\end{equation}
where of course
\begin{equation}
  \ddens_{0}(t) = \ddens_{N+1}(t) = 0
  \label{eq:PerturbationBoundary}
\end{equation}
for all $t$. 
We then plug \eqref{eq:Perturbation} into \eqref{eq:Current-Density} and~\eqref{eq:LocalDensityTimeEvo}, retaining at most terms that are linear in the perturbation, and we obtain 
\begin{equation}
  \dot{\ddens}_{n}(t) = 
  - {a}_{n} \ddens_{n}(t)
  + \dfull_{n} \ddens_{n+1}(t)
  + \oneminus{\dfull_{n}} \ddens_{n-1}(t)
  \qquad n = 1,\dots,N
  \, ,
  \label{eq:LinearizedTimeEvoMF}
\end{equation}
where
\begin{equation}
  {a}_{n} \equiv \oneminus{\dfull_{n+1}} + \dfull_{n-1}  
  + \omega_\mathrm{A} + \omega_\mathrm{D}
  \, .
  \label{eq:definizione_a}
\end{equation}
We note that the resulting system of (first-order, linear) ordinary differential equations is characterized by a coefficient matrix (the relaxation matrix) being tridiagonal. 
In the special case ${\omega_\mathrm{A} = \omega_\mathrm{D} = 0}$ (pure TASEP) we recover the relaxation matrix reported in~\cite{PelizzolaPretti17}. 
Now, a relaxation mode of the system is a solution of the type 
\begin{equation}
  \ddens_{n}(t) = {v}_{n} \rme^{-\lambda t}
  \qquad n = 0,\dots,N+1
  \, ,
  \label{eq:RelaxationMode}
\end{equation}
where $\lambda$ is the relaxation rate and, according to~\eqref{eq:PerturbationBoundary}, we have 
\begin{equation}
  {v}_{0} = {v}_{N+1} = 0
  \, .
\end{equation}
Plugging \eqref{eq:RelaxationMode} into~\eqref{eq:LinearizedTimeEvoMF} yields
\begin{equation}
  {a}_{n} {v}_{n} - \dfull_{n} {v}_{n+1} - \oneminus{\dfull_{n}} {v}_{n-1}
  = \lambda {v}_{n}
  \qquad n = 1,\dots,N
  \, ,
\end{equation}
that is, the eigenvalue problem associated with the relaxation matrix. 
Section~\ref{sec:rigorous-results} will be entirely devoted to the analysis of such a problem, focussing in particular on the smallest $\lambda$, which corresponds to the slowest relaxation mode, i.e.~the one relevant at long times. 
Here we just observe that, even though the relaxation matrix is nonsymmetric, the fact that its off-diagonal entries never change sign allows one to perform a simple similarity transformation (which preserves the eigenvalues) to a symmetric matrix. 
In formulae, defining 
\begin{equation}
  {u}_{n} \equiv {v}_{n} 
  \prod_{k=0}^{n-1}
  \sqrt{\frac{\dfull_{k}}{\oneminus{\dfull_{k+1}}}} 
  \qquad n = 0,\dots,N+1 
\end{equation}
(it is understood that the product is $1$ for ${n = 0}$), we get
\begin{equation}
  {a}_{n} {u}_{n} 
  - \sqrt{\dfull_{n} \oneminus{\dfull_{n+1}}} \, {u}_{n+1} 
  - \sqrt{\dfull_{n-1} \oneminus{\dfull_{n}}} \, {u}_{n-1}
  = \lambda {u}_{n}
  \qquad n = 1,\dots,N
\end{equation}
with the usual boundary conditions
\begin{equation}
  {u}_{0} = {u}_{N+1} = 0
  \, .
\end{equation}

%% file: TASEP-LK_phase-diagram.tex
\section{Phase diagram}
\label{sec:phase-diagram}

\begin{figure}
	\centering
	\resizebox{100mm}{!}{\includegraphics*{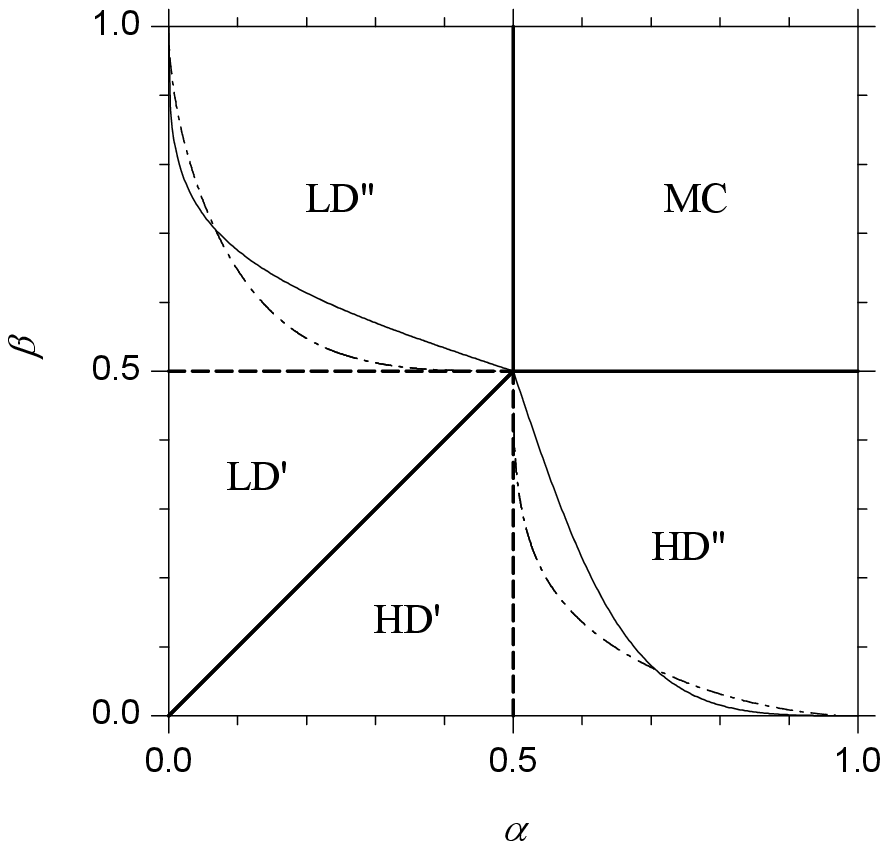}}
	\caption
	{
		Pure TASEP phase diagram. 
		Static and dynamical transitions are denoted by thick and thin lines, respectively. 
		Phase labels are explained in the text. 
		Thick dashed lines denote (static) subphase boundaries. 
		Thin dash-dotted lines denote the (approximate) dynamical transition predicted by the mean-field theory (see section~\ref{sec:dynamical-transition}).
	}
	\label{fig:phase_diagram_TASEP}
\end{figure}

The phase diagram of pure TASEP with open boundary conditions is reported in figure~\ref{fig:phase_diagram_TASEP}, where both static and dynamical transitions are shown. 
All transition lines in this phase diagram are exactly known~\cite{deGierEssler05,deGierEssler06,deGierEssler08,DerridaDomanyMukamel92,SchutzDomany93,Derrida-etal93,Derrida98}. 
One can observe three main phases, usually named high-density (HD), low-density (LD) and maximal current (MC) phases, separated by static transitions. 
These three phases differ in the values of the current ${J}$ (which is uniform in the steady state) and of the bulk density $\dbulk$ (the limiting value of the local density, far from either boundary). 

In the HD phase (${\beta < 1/2}$, ${\alpha > \beta}$), the current ${{J} = \beta \oneminus{\beta}}$ and the bulk density ${\dbulk = \oneminus{\beta} > 1/2}$ do not depend on the injection rate $\alpha$. 
From the left (injection) side, the density approaches its bulk value with an exponential decay (HD' region), which for ${\alpha \ge 1/2}$ (HD'' region) also exhibits power-law corrections. 
Note that, since the latter distinction characterizes only a finite portion of the system, it cannot be regarded as a true thermodynamic transition, so that it is usually denoted as a \emph{subphase} boundary~\cite{ProemeBlytheEvans11}.
The LD phase is closely related to the HD phase, as the TASEP is symmetric under the particle-hole transformation ${\alpha \leftrightarrow \beta}$, ${\oneminus{\dfull_{n}} \leftrightarrow \dfull_{N+1-n}}$. 
As a consequence, in the LD phase (${\alpha < 1/2}$, ${\beta > \alpha}$), the current ${J} = \alpha \oneminus{\alpha}$ and the bulk density ${\dbulk = \alpha < 1/2}$ do not depend on the extraction rate $\beta$. 
From the right (extraction) side the density approaches its bulk value with an exponential decay (LD' region), which for ${\beta \ge 1/2}$ (LD'' region) is also characterized by power-law corrections. 
The LD-HD transition is generally classified as a first-order transition, because it can be regarded as a coexistence of both phases. 
Finally, in the MC phase (${\alpha > 1/2}$, ${\beta > 1/2}$), the current ${{J} = 1/4}$ and the bulk density ${\dbulk = 1/2}$ are independent of both $\alpha$ and $\beta$, and the approach to the bulk density is power-law from both sides. 
Let us remark that the mean-field theory fulfils the model's symmetry and correctly predicts the static phase transitions (and their locations), except the distiction between the HD' and HD'' subphases (and between LD' and LD'' as well), which was first detected by the exact steady-state solution~\cite{SchutzDomany93}. 

All the above phase transitions (except the subphase ones) are also accompanied by a singular behavior of the slowest relaxation rate of the system, i.e. by \emph{dynamical transitions}. 
In particular, this relaxation rate (also named \emph{gap} in the theorists' jargon~\cite{deGierEssler05,deGierEssler06,deGierEssler08}) turns out to vanish along the HD-LD transition line and in the whole MC phase region, so that, at this level, one could state that the static and dynamical phase diagrams coincide. 
Nevertheless, a special dynamical transition, not corresponding to any static transition, has recently been discovered theoretically~\cite{deGierEssler05,deGierEssler06,deGierEssler08} and confirmed numerically~\cite{ProemeBlytheEvans11}. 
Such a transition lies within the HD'' phase region, and its symmetric  LD'' counterpart. 
Considering the HD case to fix ideas, the transition is characterized by a critical value of $\alpha$ (say $\alpha_\mathrm{c}$), depending on $\beta$ as 
\begin{equation}
  \label{eq:dGEline}
  \alpha_\mathrm{c}(\beta) = \left[ 1 + \left( \frac{\beta}{\oneminus{\beta}}
  \right)^{1/3} \right]^{-1}
  \, ,
\end{equation}
such that for ${\alpha > \alpha_\mathrm{c}(\beta)}$ the gap is independent of $\alpha$, whereas for ${\alpha < \alpha_\mathrm{c}(\beta)}$ it turns out to depend on both $\alpha$ and $\beta$. 

\begin{figure}
	\centering
	\resizebox{100mm}{!}{\includegraphics*{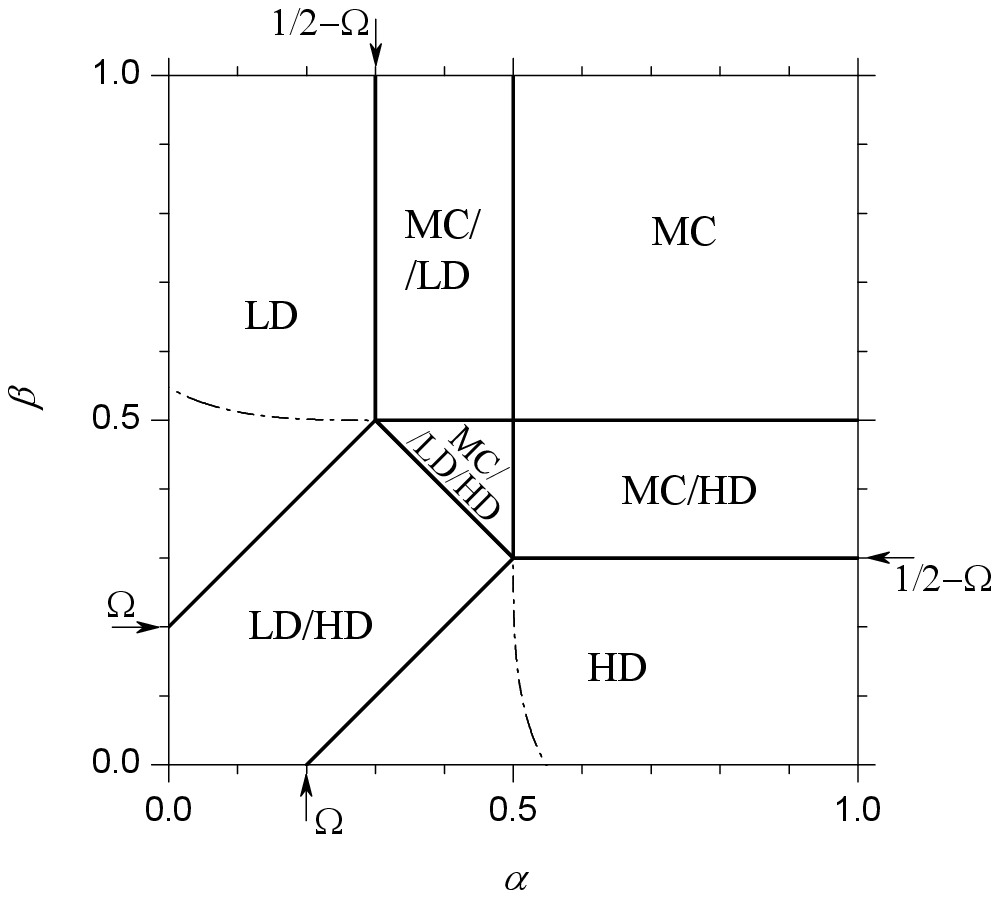}}
	\caption
	{
		Symmetric TASEP-LK phase diagram. 
		Static and dynamical transitions are denoted by thick and thin lines, as in the previous figure. 
		Phase labels are explained in the text; 
		multiple labels denote coexistence regions.
		Thin dash-dotted lines denote the (approximate) dynamical transition predicted by the mean-field theory (see section~\ref{sec:dynamical-transition}).
	}
	\label{fig:phase_diagram_TASEP-LK}
\end{figure}

A better understanding of the physical meaning of this dynamical transition was left as an open question in \cite{ProemeBlytheEvans11} and is one of the aims of the current work. 
The numerical investigation of the mean-field relaxation matrix, reported in \cite{PelizzolaPretti17}, has shown that the behavior of the slowest relaxation rate can be reproduced in a qualitatively correct way by the mean-field theory, with an approximate location of $\alpha_\mathrm{c}(\beta)$. 
Moreover, it was pointed out that the spectrum of the relaxation matrix changes qualitatively at the transition, namely, for ${\alpha > \alpha_\mathrm{c}(\beta)}$ the spectrum seems to tend, in the infinite-size limit, to a continuous, Toeplitz-like band. 
Another purpose of the current paper is to extend the mean-field results concerning the special dynamical transition (from now on simply referred to as \emph{the} dynamical transition) to the TASEP-LK model in the symmetric case ${\Omega_\mathrm{A} = \Omega_\mathrm{D} = \Omega}$. 
The static phase diagram of the latter model has been investigated, at a mean-field level, by Parmeggiani, Franosch, and Frey~\cite{ParmeggianiFranoschFrey04}, who also claim its exactness, on the basis of computer simulations, as previously proved for pure TASEP~\cite{SchutzDomany93}.\footnote{Note however that a distinction of the HD phase into the subphases HD' and HD'' (and of LD into LD' and LD'') does not appear, because, in analogy with the pure TASEP, it cannot be detected by the mean-field theory alone.} 
The most notable result of~\cite{ParmeggianiFranoschFrey04} is the onset of parameter regions where two (or even all three) of the pure phases (HD, LD, MC) coexist, being separated by stable and localized domain walls. 
The latter phenomenon entails a shrinking of the HD and LD pure-phase regions, as shown in figure~\ref{fig:phase_diagram_TASEP-LK}.
In particular, the parameter region of the HD phase is now characterized by ${\beta + \Omega < 1/2}$ and ${\alpha > \beta + \Omega}$, whereas the LD phase by analogous relations with $\alpha$ and $\beta$ exchanged. 
Note that, since $\alpha$ and $\beta$ must be positive, pure HD and LD phase regions exist only for ${\Omega < 1/2}$. 
Let us also observe that the displacement of the pure HD (resp. LD) phase boundaries can be described as a whole by the mapping ${\beta \mapsto \beta + \Omega}$ (resp. ${\alpha \mapsto \alpha + \Omega}$). We stress the fact that the latter observation is not an end in itself. Indeed, we shall see that such a mapping plays a central role in the change of several properties from the pure TASEP to the symmetric TASEP-LK.

%% file: TASEP-LK_hd-phase.tex
\subsection{Density profiles}
\label{sec:hd-phase}

\begin{figure}
  \centering
  \resizebox{100mm}{!}{\includegraphics*{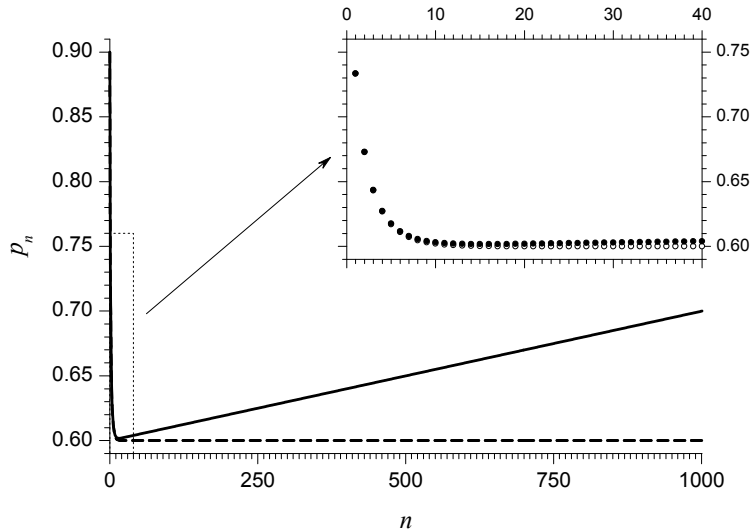}}
  \caption
  {
  	Steady-state local density $\dfull_{n}$ as a function of the node index $n$ for the TASEP-LK (${\Omega = 0.1}$, ${\beta = 0.3}$, ${\alpha = 0.9}$; solid line and solid circles) and for the pure TASEP (${\Omega = 0}$, ${\beta = 0.4}$, ${\alpha = 0.9}$; dashed line and hollow circles). 
    In both cases it turns out ${\dbulk_{0} = 0.6}$, whereas the system size is ${N = 1000}$.
  }
  \label{fig:density_profiles}
\end{figure}

Let us first study in some detail the properties of the steady-state density profile in the HD phase, that is in the parameter region defined by the following inequalities
\begin{eqnarray}
  1/2 > \Omega \geq 0
  \, , \label{eq:intervallo_valori_Omega} \\
  1/2 - \Omega > \beta > 0 
  \, , \label{eq:intervallo_valori_beta} \\
  \alpha > \beta + \Omega 
  \, . \label{eq:intervallo_valori_alpha_vs_beta}
\end{eqnarray}
In the limit of large $N$, the profile is almost completely described by a \emph{bulk solution} (being a linear function of the node index $n$ with slope $\omega$) matching the right-boundary condition~\cite{ParmeggianiFranoschFrey04}. 
The mismatch with the left-boundary condition is filled up by the onset of a \emph{boundary layer}, such that the local density approaches exponentially the bulk solution, with the characteristic length of the exponential remaining finite as $N$ grows to infinity. 
The bulk solution can be written as
\begin{equation}
  \label{eq:bulk_densities_expression}
  \dbulk_{n} = \oneminus{(\beta + \Omega)} + \omega {n}
  \qquad n = 0,\dots,N+1
  \, .
\end{equation}
For the pure TASEP one has ${\Omega = 0}$ and therefore a perfectly uniform bulk profile. 
One can easily verify that the sequence ${(\dbulk_{n})}_{n=0}^{N+1}$ satisfies both the steady-state equations \eqref{eq:SteadyStateMF} (with ${\omega_\mathrm{A} = \omega_\mathrm{D} = \omega}$) and the boundary condition~\eqref{eq:LocalDensityRight}, that is ${\dbulk_{N+1} = \oneminus{\beta}}$. 
Let us now define the quantities
\begin{equation}
  \label{eq:detrended_densities_definition}
  \ddetr_{n} \equiv \dfull_{n} - (\dbulk_{n} - \dbulk_{0})
  = \dfull_{n} - \omega {n}
  \qquad n = 0,\dots,N+1  
  \, ,
\end{equation}
which we shall call the \emph{detrended densities}, as they are the densities after subtraction of the non-uniform part of the bulk profile. 
The underlying idea is that we expect the detrended density profile to behave, at least in the HD (or LD) phase, as the density profile of some ``effective'' pure TASEP. With simple algebra one can see that, in the steady state, the detrended densities satisfy the following equations
\begin{equation}
  \label{eq:SteadyStateMF_detrended}
  \ddetr_{n} \oneminus{\ddetr_{n+1}} - \ddetr_{n-1} \oneminus{\ddetr_{n}} 
  = \omega {n} \left( \ddetr_{n+1} - \ddetr_{n-1} \right) 
  \qquad n = 1,\dots,N
  \, ,
\end{equation}
with the boundary conditions 
\begin{eqnarray}
  \label{eq:SteadyStateMF_detrended_left}
  \ddetr_{0} & = {p}_{0} & = \alpha
  \, , \\
  \label{eq:SteadyStateMF_detrended_right}
  \ddetr_{N+1} & = \dbulk_{0} & = \oneminus{(\beta+\Omega)}
  \, . 
\end{eqnarray}
Now, tentatively assuming that the detrended density profile is actually similar to that of a pure TASEP (in the HD phase), the difference ${\ddetr_{n+1} - \ddetr_{n-1}}$ would be significantly different from zero only up to finite ${n}$. 
But, in this region, the prefactor ${\omega {n}}$ vanishes, because $\omega$ is of order ${1/N}$ for ${N \to \infty}$, which leads to the conclusion that the whole right-hand side of \eqref{eq:SteadyStateMF_detrended} is almost equal to zero, and the detrended densities satisfy 
\begin{equation}
  \ddetr_{n} \oneminus{\ddetr_{n+1}} \approx \mathrm{constant}
  \qquad n = 0,1,\dots,N
  \, .
  \label{eq:SteadyStateMF_detrended_approx}
\end{equation}
The latter are in fact pure-TASEP mean-field equations (check \eqref{eq:SteadyStateMF} with ${\omega_\mathrm{A} = \omega_\mathrm{D} = 0}$), with a renormalized right-boundary condition, namely \eqref{eq:SteadyStateMF_detrended_right}. 
Thus, our assumption is justified, and the physical meaning is as follows. 
The ``rapid variations'' of the density (i.e.~those of order $1$) take place practically over a finite number of sites, where the effect of Langmuir kinetics becomes negligible for large $N$. 
This entails that the system should behave (on such a small region) as a pure TASEP with a bulk density adjusted to match the local bulk density of the TASEP-LK in that region. 
In particular, in the HD phase, the rapid variations are concentrated at the left boundary, so that the effective pure TASEP is one with a right-boundary condition renormalized to the \emph{left}-boundary value of the bulk solution \eqref{eq:bulk_densities_expression}, that is ${\dbulk_{0} = \oneminus{(\beta+\Omega)}}$.
The situation is better explained in figure~\ref{fig:density_profiles}, where we report two different stationary density profiles, one for the pure TASEP and one for the TASEP-LK, computed numerically by \eqref{eq:SteadyStateMF_iteration}, with the same $\alpha$ value, but \emph{two different} $\beta$ values, adjusted in order to obtain the same $\dbulk_{0}$.
We can actually verify that, in the vicinity of the boundary layer, the two profiles are almost indistinguishable, whereas the effect of Langmuir kinetics can be clearly appreciated at a macroscopic length scale, resulting there in a linear density profile.

The above arguments can be made rigorous as explained below. 
We study an infinite sequence ${(\ddlim_{n})}_{n=0}^{\infty}$, being a solution of \eqref{eq:SteadyStateMF_detrended_approx} taken as an equality, namely 
\begin{equation}
  \ddlim_{n} \oneminus{\ddlim_{n+1}}
  = \dbulk_{0} \oneminus{\dbulk_{0}}
  \qquad n=0,1,2,\dots
  \, ,
  \label{eq:proposizione_eta}
\end{equation}
where the constant (right-hand side) takes into account that in the bulk we expect ${\ddlim_{n} \approx \dbulk_{0}}$, and the left-boundary condition is chosen according to \eqref{eq:SteadyStateMF_detrended_left}, that is 
\begin{equation}
  \ddlim_{0} = \alpha
  \, .
  \label{eq:proposizione_eta_0}
\end{equation}
We thus have a discrete \emph{Riccati equation} \cite{Agarwal00}, whose solution can be easily worked out. 
The idea is that, in the limit of large $N$, for ${n=0,\dots,N+1}$ we will have ${\ddlim_{n} \approx \ddetr_{n}}$. 
In rigorous terms, we can put forward the following statements:
Lemma~\ref{lem:eta} determines the analytical properties of the sequence $\ddlim_{n}$ (including the simple proportionality relationship with $\sigma_n$, defined as in~\eqref{eq:definizione_sigma_ricorsiva}), whereas 
Theorem~\ref{teo:delta} establishes precise bounds for the distance between $\ddlim_{n}$ and the detrended densities $\ddetr_{n}$. 
Note that in the following we shall always consider $\dbulk_{0}$ as an alias for $\oneminus
{(\beta+\Omega)}$ (i.e. as an alternative parameter), so that the parameter bounds \eqref{eq:intervallo_valori_beta} and \eqref{eq:intervallo_valori_alpha_vs_beta} will be respectively replaced by 
\begin{eqnarray}
  1 - \Omega > \dbulk_{0} > 1/2
  \, , \label{eq:intervallo_valori_q0} \\
  \alpha > \oneminus{\dbulk_{0}} 
  \, . \label{eq:intervallo_valori_alpha}
\end{eqnarray}
\begin{lem}
  \label{lem:eta}
  Let ${(\ddlim_{n})}_{n=0}^{\infty}$ be the infinite sequence defined as follows 
  \begin{equation}
    \ddlim_{n} \equiv \dbulk_{0} 
    + (\dbulk_{0} - \oneminus{\dbulk_{0}}) 
    \left[ \left( 1 - 
    \frac{\alpha - \dbulk_{0}\hphantom{'}}{\alpha - \oneminus{\dbulk_{0}}}
    \, \gamma^{n} \right)^{-1} - 1 \right]
    \quad n = 0,1,2,\dots
    \, ,
    \label{eq:definizione_s}
  \end{equation}  
  where 
  \begin{equation}
    \gamma \equiv \frac{\oneminus{\dbulk_{0}}}{\dbulk_{0}}
    \, .
    \label{eq:definizione_gamma}
  \end{equation} 
  Then the following statements hold. 
  \begin{enumerate} 
    \item The sequence ${(\ddlim_{n})}_{n=0}^{\infty}$ satisfies \eqref{eq:proposizione_eta}, \eqref{eq:proposizione_eta_0} and 
    \begin{equation}
      \lim_{n \to \infty} \ddlim_{n} = \dbulk_{0}
      \, .
      \label{eq:limite_s}
    \end{equation}
    \item If ${(\sigma_n)}_{n=0}^{\infty}$ is the sequence defined in \eqref{eq:definizione_sigma_ricorsiva}, then   
    \begin{equation}
      \ddlim_{n} = \sigma_{n} \sqrt{\dbulk_{0} \oneminus{\dbulk_{0}}}
      \qquad n=0,1,2,\dots
      \, .
      \label{eq:definizione_sigma_vecchia}
    \end{equation}
    \item If ${\alpha \geq \dbulk_{0}}$ (resp. ${\alpha \leq \dbulk_{0}}$), then both sequences ${(\ddlim_{n})}_{n=0}^{\infty}$ and ${(\sigma_n)}_{n=0}^{\infty}$ are non-increasing (resp.~non-decreasing). 
  \end{enumerate} 
\end{lem}
The proof of Lemma~\ref{lem:eta} is trivial, so that it is omitted. 
A key-point is that the parameter bounds \eqref{eq:intervallo_valori_Omega} and \eqref{eq:intervallo_valori_q0} entail
\begin{equation}
  0 < \gamma < 1
  \, ,
  \label{eq:intervallo_valori_gamma}
\end{equation}
whereas \eqref{eq:definizione_sigma_vecchia} descends from \eqref{eq:definizione_xo} along with ${\dbulk_0 = \oneminus{(\beta + \Omega)}}$. 
\begin{teo}
  \label{teo:delta}
  Let ${(\ddetr_{n})}_{n=0}^{N+1}$ be the sequence of detrended densities, defined above, with boundary values ${\ddetr_{0} = \alpha}$ and ${\ddetr_{N+1} = \dbulk_{0}}$, and let ${(\ddlim_{n})}_{n=0}^{\infty}$ and $\gamma$ be defined according to Lemma~\ref{lem:eta}.
  Then there exist suitable positive constants $C$\footnote{Here and in the following we shall mean by \emph{constant} a finite real number, independent of the node label $n$ and of the system size ${N}$, but possibly dependent on the model parameters. Moreover, for every occurrence of the symbol $C$, it will be understood that there exists a \emph{suitable positive constant} (in principle a different one for each occurrence) verifying the relation where the symbol appears.} such that:
  \begin{equation}
    \left| \ddlim_{n} - \ddetr_{n} \right| 
    \leq \cases{ 
      C \gamma^{N}                   & \textit{if} ${\Omega = 0}$ \\
      C \frac{{n}^{2} \gamma^{n}}{N} & \textit{if} ${\Omega > 0}$ 
    } \qquad n = 0,\dots,N+1 
    \, . 
    \label{eq:upperbound_delta_modulo} 
  \end{equation}
\end{teo}
The proof of Theorem~\ref{teo:delta} is given in \ref{app:Rho} in full detail. 
In the end, we can state that the stationary density profile of the HD phase is
\begin{equation}
  {p}_{n} =  \ddetr_{n} + \omega {n}
  \approx \ddlim_{n} + \omega {n}
  \qquad n = 0,\dots,N+1
  \, ,
\end{equation}
where Lemma~\ref{lem:eta} supplies the analytical form of $\ddlim_{n}$ and Theorem~\ref{teo:delta} controls the approximation. 
It is also possible to show that Theorem~\ref{teo:delta} entails relevant consequences regarding the steady-state current profile, in particular that the latter is very close (for large ${N}$) to the current profile corresponding to the bulk solution alone (\emph{bulk current}), i.e.
\begin{equation}
  {J}_{n} = \dfull_{n} \oneminus{\dfull_{n+1}}
  \approx \dbulk_{n} \oneminus{\dbulk_{n+1}}
  \qquad n = 0,\dots,N
  \, .
\end{equation}
The latter result implies in turn that the maximum current value is very close to the bulk current at the left boundary. In rigorous terms, we can state the following.
\begin{cor}
	\label{cor:current}
	Let ${({J}_{n})}_{n=0}^{N}$ and ${(\dbulk_{n})}_{n=0}^{N+1}$ be respectively the sequences of (steady-state) currents and bulk densities, defined above, and let $\gamma$ be defined according to Lemma~\ref{lem:eta}. 
	Then, the following statements hold:	
	\begin{equation}
	  | {J}_{n} - \dbulk_{n} \oneminus{\dbulk_{n+1}} | 
	  \leq \cases{ 
		C \gamma^{N}                 & \textit{if} ${\Omega = 0}$ \\
	    C \frac{(n+1) \gamma^{n}}{N} & \textit{if} ${\Omega > 0}$
	  } \quad \ n = 0,\dots,N
	  \, , 
	  \label{eq:bound_current}
	\end{equation}
	\begin{equation}
	  \left| 
	    \max {({J}_{n})}_{n=0}^{N} 
	    - \dbulk_{0}\oneminus{\dbulk_{0}}
	  \right| 
	  \leq \cases{ 
		C \gamma^{N}  & \textit{if} ${\Omega = 0}$ \\
		C {N}^{-1}    & \textit{if} ${\Omega > 0}$
	  } 
	  \, .
	  \label{eq:bound_current-max}
	\end{equation}
\end{cor}

%% file: TASEP-LK_eigenvalues.tex
\subsection{Relaxation rates}
\label{sec:relaxation-rates}

In section~\ref{sec:model} we have seen that the relaxation rates coincide with the eigenvalues of a tridiagonal symmetric matrix, which we now call $A$, whose associated linear operator maps the vector ${u \equiv (u_1,\dots,u_N)}$ to ${Au \equiv (Au_1,\dots,Au_N)}$, defined componentwise as
\begin{equation}
  {A}{u}_{n} \equiv (\oneminus{\ddetr_{n+1}} + \ddetr_{n-1}) \, {u}_{n} 
  - \sqrt{\dfull_{n} \oneminus{\dfull_{n+1}}} \, {u}_{n+1}  
  - \sqrt{\dfull_{n-1} \oneminus{\dfull_{n}}} \, {u}_{n-1}  
  \, ,
  \label{eq:definizione_A}
\end{equation}
where ${{u}_{0} \equiv {u}_{N+1} \equiv 0}$. 
Note that we have taken into account \eqref{eq:detrended_densities_definition} in combination with \eqref{eq:definizione_a} to get a simpler expression of the diagonal terms. 
The slowest relaxation rate is the smallest eigenvalue, which here we shall simply denote by $\lambda$. 
The aim of this section is a detailed investigation about the limit value of $\lambda$ for ${N \to \infty}$, including the determination of bounds for its scaling behavior at large $N$. 

First of all, it is useful to discuss a preliminary estimate of $\lambda$ in terms of the smallest eigenvalue $\mu$ of a modified matrix $B$ (whose associated linear operator maps ${u \equiv (u_1,\dots,u_N)}$ to ${Bu \equiv (Bu_1,\dots,Bu_N)}$), defined by
\begin{equation}
  {B}{u}_{n} \equiv (\oneminus{\ddlim_{n+1}} + \ddlim_{n-1}) \, {u}_{n} 
  - \sqrt{\dbulk_{n} \oneminus{\dbulk_{n+1}}} \, {u}_{n+1}  
  - \sqrt{\dbulk_{n-1} \oneminus{\dbulk_{n}}} \, {u}_{n-1} 
  \, ,
  \label{eq:definizione_B}
\end{equation}
still with ${{u}_{0} \equiv {u}_{N+1} \equiv 0}$. 
The matrix $B$ is symmetric as well as $A$, which guarantees that its eigenvalues are still all real. 
Moreover, the change from $A$ to $B$ is small, as it consists in replacing the detrended densities $\ddetr_{n}$ with their infinite-size approximation $\ddlim_{n}$ (in the diagonal terms) and the currents $\dfull_{n}\oneminus{\dfull_{n+1}}$ with the corresponding bulk currents $\dbulk_{n}\oneminus{\dbulk_{n+1}}$ (in the off-diagonal terms). 
Precise bounds for the distances between these quantities have been previously stated in Theorem~\ref{teo:delta} and Corollary~\ref{cor:current}.
One then expects that also the eigenvalues should be moderately affected by this change. 
As far as the smallest eigenvalue is concerned, we can prove the following result, which is in fact a direct consequence of Theorem~\ref{teo:delta} and Corollary~\ref{cor:current}.
\begin{cor}
  \label{cor:simplified-matrix}
  Let $\lambda$ and $\mu$ be the smallest eigenvalues of matrices ${A}$ and ${B}$, respectively, and let $\gamma$ be defined according to Lemma~\ref{lem:eta}.
  Then
  \begin{equation}
	| \lambda - \mu | \leq \cases{ 
	  C \gamma^{N} & \textit{if} ${\Omega = 0}$ \\
	  C {N}^{-1}   & \textit{if} ${\Omega > 0}$ 
    }
	\, .
	\label{eq:upperbound_lambda-mu}
  \end{equation}
\end{cor}
Note that, on the one hand, the role of the above result is technical, since the eigenvalue problem for matrix $B$ turns out to be simpler than for matrix $A$. 
On the other hand, the structure of matrix $B$, along with the related results for the smallest eigenvalue, reveals that the dynamical transition is driven by the competition between the boundary layer (diagonal terms) and the bulk (off-diagonal terms). 
In the special case of pure TASEP the bulk is uniform, so that the only deviation from the Toeplitz structure arises from the diagonal terms, as observed in \cite{PelizzolaPretti17}.

From the technical point of view, Corollary~\ref{cor:simplified-matrix} suggests to focus on $\mu$ in order to investigate the leading behavior of $\lambda$ at large $N$. 
The basic idea is to find suitable upper- and lower-bounds, which may tend to coincide in the limit ${N \to \infty}$. 
The \emph{Courant minimax principle} allows one to obtain a multitude of upper-bounds in the form ${\mu \leq (u,Bu)}$, for any vector ${u \in \mathbb{R}^{N}}$ such that ${\|{u}\| = 1}$, where ${(u,v) \equiv \sum_{n=1}^{N} {u}_{n} {v}_{n}}$ is the usual Euclidean scalar product and ${\|{u}\| \equiv \sqrt{(u,u)}}$ the corresponding norm. 
Thus, taking into account \eqref{eq:definizione_B} and \eqref{eq:definizione_sigma_vecchia}, we can immediately prove the following. 
\begin{lem}
  \label{lem:Courant-type_bound}
  Let $\mu$ be the smallest eigenvalue of matrix ${B}$ and let $\sigma_n$ be defined by \eqref{eq:definizione_sigma_ricorsiva}. 
  Let ${u_1,\dots,u_N}$ be real numbers such that ${\sum_{n=1}^{N} {{u}_{n}}^{2} = 1}$. 
  Then
  \begin{equation}
    \mu \leq 1 
    - \sqrt{\dbulk_{0} \oneminus{\dbulk_{0}}} \, 
    \sum_{n=1}^{N} \left( \sigma_{n+1} - \sigma_{n-1} \right) {{u}_{n}}^2
    - 2 \sum_{n=1}^{N-1} \sqrt{\dbulk_{n} \oneminus{\dbulk_{n+1}}} \, {u}_{n} {u}_{n+1}
    \, .
  \label{eq:Courant-type_bound}
  \end{equation}
\end{lem}
Regarding lower-bounds, the right way to proceed is less obvious. We face the problem in the same spirit as the \emph{Gershgorin circle theorem}'s, which allows us to formulate the following (see \ref{app:Gershgorin-type_bound}). 
\begin{lem}
  \label{lem:Gershgorin-type_bound}
  Let $\mu$ be the smallest eigenvalue of matrix ${B}$ and let $\sigma_n$ be defined by \eqref{eq:definizione_sigma_ricorsiva}. 
  Let ${w_{0},\dots,w_{N+1}}$ be real positive numbers (except $w_{0}$, possibly being zero). Then
  \begin{equation}
    \mu \geq 1 - \sqrt{\dbulk_{0} \oneminus{\dbulk_{0}}} \, \max \left(
      \sigma_{n+1} - \sigma_{n-1} + \frac{{w}_{n+1} + {w}_{n-1}}{{w}_{n}} 
    \right)_{n = 1}^{N} 
    \, .
    \label{eq:Gershgorin-type_bound}
  \end{equation}
\end{lem}

We are now in a position to apply the strategy outlined above. 
Let us note in advance that the difficulties one encounters in such a task are dramatically different, depending on the boundary conditions for the detrended densities, which we recall to be ${\ddetr_{0} = \alpha}$ (left) and ${\ddetr_{N+1} = \dbulk_{0}}$ (right), according respectively to \eqref{eq:SteadyStateMF_detrended_left} and \eqref{eq:SteadyStateMF_detrended_right}. 
In particular, the case ${\alpha \geq \dbulk_{0}}$ turns out to be much simpler than the complementary one ${\alpha < \dbulk_{0}}$. 
This is ultimately related to the fact that, according to Lemma~\ref{lem:eta}, in the former case the sequence ${(\sigma_{n})}_{n=0}^{\infty}$ is non-increasing, so that ${\sigma_{n+1} - \sigma_{n-1}}$ can never be positive. 
In such a case, Lemma~\ref{lem:Gershgorin-type_bound} with the simple choice ${{w}_{n} \equiv 1}$ for all $n$ (which can be shown to coincide with the usual Gershgorin theorem) immediately yields the following result. 
\begin{lem}
  \label{lem:lowerbound_mu_alfa_grandi}
  Let $\mu$ be the smallest eigenvalue of matrix ${B}$. 
  If ${\alpha \geq \dbulk_{0}}$, then 
  \begin{equation}
    \mu \geq 1 - 2 \sqrt{\dbulk_{0} \oneminus{\dbulk_{0}}} 
    \, .
  \end{equation}
\end{lem}
On the other hand, regarding upper-bounds, a suitable choice of ${u_1,\dots,u_N}$ in Lemma~\ref{lem:Courant-type_bound} allows us to prove the following (see \ref{app:upperbound_mu}).
\begin{lem}
  \label{lem:upperbound_mu}
  Let $\mu$ be the smallest eigenvalue of matrix ${B}$. Then
  \begin{equation}
	\mu \leq 1 - 2 \sqrt{\dbulk_{0} \oneminus{\dbulk_{0}}}  
	+ \cases{ 
		C {N}^{-2}   & \textit{if} ${\Omega = 0}$ \\
		C {N}^{-2/3} & \textit{if} ${\Omega > 0}$
	}
	\, .
  \end{equation}
\end{lem}
We shall see that, when ${\alpha < \dbulk_{0}}$, the latter bound may no longer be a good one, even though, at odds with Lemma~\ref{lem:lowerbound_mu_alfa_grandi}, it holds in principle with no restriction on $\alpha$ (except ${\alpha > \oneminus{\dbulk_0}}$, which is required in order to stay within the HD-phase region, as it is always understood). 

Let us now switch to consider the case ${\alpha < \dbulk_{0}}$, which, as previously mentioned, requires considerably more effort. 
The basic idea is that, in order to obtain good bounds, we have to choose ${u}_{n}$ in Lemma~\ref{lem:Courant-type_bound} (Courant-type bound) and ${w}_{n}$ in Lemma~\ref{lem:Gershgorin-type_bound} (Gershgorin-type bound) as close as possible to the actual eigenvector. 
Unfortunately, the eigenvalue problem for matrix ${B}$ is still too difficult to be analyzed directly, especially because it still depends on the size $N$, so that one has to devise some alternative route. 
Recalling the bulk density expression \eqref{eq:bulk_densities_expression}, namely ${\dbulk_{n} = \dbulk_{0} + \omega{n}}$, we notice that, if ${n \ll N}$, we have ${\dbulk_{n} \approx \dbulk_{0}}$ (and thence also ${\dbulk_{n} \oneminus{\dbulk_{n+1}} \approx \dbulk_{0} \oneminus{\dbulk_{0}}}$) independently of ${n}$.
In other words, if we observe the matrix ${B}$ upon increasing $N$, but only up to row and column indices remaining much smaller than $N$, the off-diagonal entries tend to be constant, approaching the value $- \sqrt{\dbulk_{0} \oneminus{\dbulk_{0}}}$.
As a consequence, from \eqref{eq:definizione_B} in combination with \eqref{eq:definizione_sigma_vecchia} we expect that, for large $N$ and ${n \ll N}$, the eigenvector components ${v}_{n}$ satisfy the following simple equation 
\begin{equation}
  \left( \sigma_{n+1} - \sigma_{n-1} \right) {v}_{n} + {v}_{n+1} + {v}_{n-1} 
  \propto {v}_{n} 
  \, ,
  \label{eq:successione_v_euristica}
\end{equation}
with $\sigma_{n}$ defined in \eqref{eq:definizione_sigma_ricorsiva} and ${{v}_{0} = 0}$. 
Note that the above equation no longer depends on $N$, and indeed we expect it to become more and more accurate as $N$ grows large, even though only as far as ${n \ll N}$.
Now, the idea is to study the properties of a sequence ${({v}_{n})}_{n=0}^{\infty}$ satisfying \eqref{eq:successione_v_euristica}, as a function of the proportionality coefficient. 
Such a sequence can be defined by recursion, with the ``initial'' conditions ${{v}_{0} = 0}$ (by hypothesis) and an arbitrary ${v}_{1}$ (for instance ${{v}_{1} = 1}$). 
In particular, we are interested in studying the hypotheses under which the sequence never changes sign (so that it may be employed in the Gershgorin-type bounds) and whether it goes to zero rapidly enough (so that it may be of use in the Courant-type bounds).
In the end, the relevant properties can be formally stated as follows.
\begin{lem}
  \label{lem:successione_v}
  Let ${(\sigma_{n})}_{n=0}^{\infty}$ be the sequence defined by \eqref{eq:definizione_sigma_ricorsiva}, in the hypothesis ${\alpha < \dbulk_{0}}$, and let ${({v}_{n}(x))}_{n=0}^{\infty}$ be the family of sequences defined by \eqref{eq:definizione_v_cond_iniz} and \eqref{eq:definizione_v}, parameterized by ${x \in \mathbb{R}}$. \\ 
  Then the following statements hold. 
  \begin{enumerate}
  	\item \label{statement1} The sequence ${({v}_{n}(x))}_{n=0}^{\infty}$ is nonoscillatory (i.e.~eventually positive or eventually negative) if and only if ${x \geq 1}$. 
	\item \label{statement2} The set $\mathcal{X}$ of real numbers ${x}$ such that ${{v}_{n}(x) > 0}$ for all ${n>0}$ is a closed, infinite interval $\mathcal{X} = [{x}_{*},\infty)$, where ${{x}_{*} \equiv \inf \mathcal{X} \in [1,{x}_{\circ})}$, ${x}_{\circ}$ being defined by \eqref{eq:definizione_xo}. 
	\item \label{statement3} Let ${\zeta: [1,\infty) \to (0,1]}$ be the function defined by \eqref{eq:definizione_z} and let ${{f}_{n}: [1,\infty) \to \mathbb{R}}$ be the functions defined as 
    \begin{equation}
      f_n(x) \equiv \sum_{k=0}^{n} \left( \sigma_{k+1} - \sigma_{k-1} \right) 
      v_k(x) {\zeta(x)}^{k}
      \qquad n=0,1,2,\dots
      \label{eq:definizione_fnx}
    \end{equation}
    (with an arbitrary definition of $\sigma_{-1}$).
	Then, the limit 
	\begin{equation}
	  {f}({x}) \equiv \lim_{n \to \infty} {f}_{n}(x)
	  \label{eq:definizione_fx}
	\end{equation}
	exists and is finite for all ${x \geq 1}$. 
	Moreover, if the sequence ${({v}_{n}(x))}_{n=0}^{\infty}$ is eventually positive (resp.~eventually negative), then ${{f}({x}) \leq 1}$ (resp. ${{f}({x}) \geq 1}$). 
	\item \label{statement4} If ${{x}_{*} > 1}$, then ${{f}({x}_{*})=1}$. 
	\item \label{statement5} If ${{x}_{*} > 1}$, then ${{v}_{n}({x}_{*}) \leq {C}{\zeta({x}_{*})}^{n}}$ for all ${n}$.
  \end{enumerate}
\end{lem} 
As mentioned above, these results have important consequences in terms of bounds. 
In particular, statement \eqref{statement2} allows us to choose ${{w}_{n} = {v}_{n}(x)}$ in the Gershgorin-type bound (Lemma~\ref{lem:Gershgorin-type_bound}), for all ${{x} \in \mathcal{X}}$. 
One can easily realize that the most restrictive bound is attained for the smallest ${x}$ value, which immediately leads to the following.
\begin{lem}
  \label{lem:lowerbound_mu_alfa_piccoli}
  Let $\mu$ be the smallest eigenvalue of matrix ${B}$, and let ${x}_{*}$ be defined as in Lemma~\ref{lem:successione_v}. 
  If ${\alpha < \dbulk_{0}}$, then 
  \begin{equation}
    \mu \geq 1 - 2 {x}_{*} \sqrt{\dbulk_{0} \oneminus{\dbulk_{0}}} 
    \, .
  \end{equation}
\end{lem}
On the other hand, regarding upper-bounds, the next result can be proved by slightly more complicated manipulations, which make use in particular of statement (v), as detailed in \ref{app:upperbound_mu_alfa_piccoli}. 
\begin{lem}
  \label{lem:upperbound_mu_alfa_piccoli}
  Let $\mu$ be the smallest eigenvalue of matrix ${B}$, and let $\zeta({x})$ and ${x}_{*}$ be defined as in Lemma~\ref{lem:successione_v}. 
  If ${\alpha < \dbulk_{0}}$ and ${{x}_{*} > 1}$, then
  \begin{equation}
    \mu \leq 1 - 2 {x}_{*} \sqrt{\dbulk_{0} \oneminus{\dbulk_{0}}}  
    + \cases{ 
  	  C \zeta({x}_{*})^{2{N}}  & \textit{if} ${\Omega = 0}$ \\
  	  C {N}^{-1}               & \textit{if} ${\Omega > 0}$
    }
    \, .
  \end{equation}
\end{lem}
As previously mentioned, in the regime ${\alpha < \dbulk_{0}}$ the bound stated by Lemma~\ref{lem:upperbound_mu} still holds but, as soon as $\alpha$ drops below the dynamical transition (so that ${x_* > 1}$), this is no longer a good bound, as Lemma~\ref{lem:upperbound_mu_alfa_piccoli} turns out to be stronger. 
This is the parameter region where the (infinite-size) relaxation rate begins to depend on $\alpha$ and gets lower than its plateau value.

Let us now finally get back to $\lambda$, which is the physically important quantity. 
Making use of the results proved so far, we can state the following.
\begin{teo}
  \label{teo:lambda}
  Let $\lambda$ be the smallest eigenvalue of matrix ${A}$, and let $\zeta({x})$ and ${x}_{*}$ be defined as in Lemma~\ref{lem:successione_v}. 
  Either if ${\alpha \geq \dbulk_{0}}$, or if ${\alpha < \dbulk_{0}}$ with ${{x}_{*} = 1}$, then
  \begin{equation}
    \left| \lambda - 1 + 2 \sqrt{\dbulk_{0} \oneminus{\dbulk_{0}}} \right| 
    \leq \cases{ 
  	  C {N}^{-2}   & \textit{if} ${\Omega = 0}$ \\
  	  C {N}^{-2/3} & \textit{if} ${\Omega > 0}$
    }
    \, .
  \end{equation}
  Otherwise, if ${\alpha < \dbulk_{0}}$ with ${{x}_{*} > 1}$, then
  \begin{equation}
    \left| \lambda - 1 + 2 {x}_{*} \sqrt{\dbulk_{0} \oneminus{\dbulk_{0}}} \right| 
    \leq \cases{ 
  	  C \zeta({x}_{*})^{2{N}}  & \textit{if} ${\Omega = 0}$ \\
  	  C {N}^{-1}               & \textit{if} ${\Omega > 0}$
    }
    \, .
  \end{equation}
\end{teo}
The proof of this last theorem consists of two parts. 
First, by Lemmas \ref{lem:lowerbound_mu_alfa_grandi}, \ref{lem:upperbound_mu}, \ref{lem:lowerbound_mu_alfa_piccoli} and~\ref{lem:upperbound_mu_alfa_piccoli} all together, an analogous theorem is immediately proved for $\mu$ (rather than $\lambda$). 
Second, Corollary~\ref{cor:simplified-matrix} allows one to extend the result from $\mu$ to $\lambda$, due to the fact that in any case the distance between $\mu$ and $\lambda$ is, for ${N \to \infty}$, infinitesimal of a higher order than the distance between $\lambda$ and its limit value (or at most of the same order). 
Let us note that, as far as the pure TASEP is concerned (${\Omega = 0}$), the latter step requires the condition ${\zeta({x}_{*})^{2} \geq \gamma}$, which is proved below. 
Using \eqref{eq:definizione_gamma} and \eqref{eq:definizione_xo} with ${\beta + \Omega = \oneminus{\dbulk_{0}}}$, simple algebra shows that 
\begin{equation}
  \gamma = {\zeta(x_\circ)}^{2}
  \, , 
  \label{eq:relazione_gamma-zetaxo}
\end{equation}
so that (noting that ${\zeta(x) > 0}$ for all ${x \geq 1}$), the desired inequality becomes ${\zeta({x}_{*}) \geq \zeta({x}_{\circ})}$. 
Now, since $\zeta(x)$ is monotonically decreasing, the latter condition is satisfied by ${{x}_{*} \leq {x}_{\circ}}$, which turns out to be guaranteed by Lemma~\ref{lem:successione_v}, statement \eqref{statement2}.

Let us finally remark that, in order to prove Theorem~\ref{teo:lambda}, statements \eqref{statement1}, \eqref{statement3} and \eqref{statement4} in Lemma~\ref{lem:successione_v} are apparently unnecessary. 
In fact statement \eqref{statement1} is a basic step in the proof of Lemma~\ref{lem:successione_v} itself, whereas \eqref{statement3} and \eqref{statement4} provide analytical tools for investigating the behavior of ${x}_{*}$ as a function of the model parameters, and thence for detecting the dynamical transition. 
In particular, assuming the same definitions of $x_*$, $x_\circ$ and $f(x)$ as in Lemma~\ref{lem:successione_v}, $f(x)$ being obviously the same as in \eqref{eq:definizione_fx_come_serie}, we can formulate the following. 
\begin{cri}
  If equation ${f(x) = 1}$ has no solution ${x \in (1,x_\circ)}$, then ${x_* = 1}$. 
  \label{cri:uno}
\end{cri}
\begin{cri}
  If there exists ${x \geq 1}$ such that ${f(x) > 1}$, then ${x_* > x}$ and ${f(x_*) = 1}$. 
  \label{cri:due}
\end{cri}
Criterion~\ref{cri:uno} is an immediate consequence of statement~\eqref{statement4} and of the fact that, according to \eqref{statement2}, ${x_* \in [1,x_\circ)}$. 
In the discussion of section~\ref{sec:dynamical-transition}, this criterion is used to determine the parameter range where ${x_* = 1}$. 
Criterion~\ref{cri:due} can be proved observing that, according to statements \eqref{statement2} and~\eqref{statement3}, the condition ${x \geq x_*}$ implies ${x \in \mathcal{X}}$ and thence ${f(x) \leq 1}$. 
Therefore, by complementarity, ${f(x) > 1}$ implies ${x < x_*}$. 
Since ${x \geq 1}$ by hypothesis, the fact that ${f(x_*) = 1}$ follows immediately from statement~\eqref{statement4}. 
Let us note that, in the special case ${x = 1}$, this criterion states that ${f(1) > 1}$ implies ${x_* > 1}$. 
This fact is used in section~\ref{sec:dynamical-transition} to determine the parameter range where the latter condition holds, whereas the precise $x_*$ value is determined by solving (numerically) ${f(x_*) = 1}$. 
Moreover, as for certain parameter values this last equation turns out to have two solutions, Criterion~\ref{cri:due} allows us to rule out the spurious one, say $\tilde{x}_*$, because in such cases we observe the occurrence of a range of $x$ values larger than $\tilde{x}_*$ verifying ${f(x) > 1}$.

%% file: TASEP-LK_appendix-C.tex
\section{Existence and uniqueness of the stationary solution}
\label{app:EU}

In this section we state and prove the existence and uniqueness results for the stationary local densities $\dfull_{n}$, which we have introduced in section~\ref{sec:model}. 
In particular, Lemma~\ref{lem:noboundary} shows that, in a solution being physically meaningful (i.e. such that ${\dfull_{n} \in [0,1]}$ for all ${n = 1,\dots,N}$), the local densities can never be $0$ or $1$. 
This is a useful preliminary step to prove Lemma~\ref{lem:existence} (existence) and Lemma~\ref{lem:uniqueness} (uniqueness), and is also invoked in the proof of Lemma~\ref{lem:r_bounds} (\ref{app:Rho}). 
Note that we keep the possibility of asymmetry (${\omega_\mathrm{A} \neq \omega_\mathrm{D}}$) but, for reasons that will be clear in the proof, Lemma~\ref{lem:existence} leaves out the pure TASEP case (${\omega_\mathrm{A} = \omega_\mathrm{D} = 0}$). 
Theorem~\ref{teo:EU} fills this gap and provides the complete result. 
Throughout the remainder of this section, we shall always take the following hypotheses: 
$N$ is a positive integer, $\alpha$ and $\beta$ are positive real numbers, whereas $\omega_\mathrm{A}$ and $\omega_\mathrm{D}$ are non-negative real numbers.

\begin{lem}
  \label{lem:noboundary}
  If the $N$-tuple ${(\dfull_{1},\dots,\dfull_{N}) \in {[0,1]}^{N}}$ is a solution of the stationary-state equation \eqref{eq:SteadyStateMF} with the boundary conditions \eqref{eq:LocalDensityLeft} and \eqref{eq:LocalDensityRight}, then ${(\dfull_{1},\dots,\dfull_{N}) \in {(0,1)}^{N}}$. 
\end{lem}

\paragraph{Proof}  We have to prove that ${\dfull_{k} \neq 0}$ and ${\dfull_{k} \neq 1}$ for any ${k=1,\dots,N}$. 
By contradiction, let us assume ${\dfull_{n} = 0}$ for a certain ${n \in \{1,\dots,N\}}$. 
Then from \eqref{eq:SteadyStateMF} we immediately get ${\dfull_{n-1} = -\omega_\mathrm{A} \leq 0}$. 
Now, if ${n = 1}$, we have ${\dfull_{0} \leq 0}$, which is a contradiction, since the boundary condition \eqref{eq:LocalDensityLeft} imposes ${\dfull_{0} = \alpha > 0}$.
Otherwise, if ${n > 1}$, we have ${\dfull_{n-1} \leq 0}$, which means ${\dfull_{n-1} = 0}$, since ${\dfull_{k} \geq 0}$ for all ${k=1,\dots,N}$ by hypothesis. 
We can then proceed by induction, still obtaining ${\dfull_{0} \leq 0}$, which we have already shown to be a contradiction. 
The proof for ${\dfull_{k} \neq 1}$ is fully analogous, as we see below. 
Let us assume ${\dfull_{n} = 1}$ for a certain ${n \in \{1,\dots,N\}}$. 
Then from \eqref{eq:SteadyStateMF} we immediately get ${\oneminus{\dfull_{n+1}} = -\omega_\mathrm{D} \leq 0}$. 
Now, if ${n = N}$, we have ${\dfull_{N+1} \geq 1}$, which is a contradiction, since the boundary condition \eqref{eq:LocalDensityRight} imposes ${\dfull_{N+1} = \oneminus{\beta} < 1}$.
Otherwise, if ${n < N}$, we have ${\dfull_{n+1} \geq 1}$, which means ${\dfull_{n+1} = 1}$, since ${\dfull_{k} \leq 1}$ for all ${k=1,\dots,N}$ by hypothesis. 
We can then proceed by induction, still obtaining ${\dfull_{N+1} \geq 1}$, which we have already shown to be a contradiction.

\begin{lem}
	\label{lem:existence}
	If ${\omega_\mathrm{A} + \omega_\mathrm{D} > 0}$, then the stationary-state equation \eqref{eq:SteadyStateMF} with the boundary conditions \eqref{eq:LocalDensityLeft} and \eqref{eq:LocalDensityRight} admits a solution $(\dfull_{1},\dots,\dfull_{N}) \in {[0,1]}^{N}$. 
\end{lem}

\paragraph{Proof}  As we are only interested in solutions $(\dfull_{1},\dots,\dfull_{N}) \in {[0,1]}^{N}$, Lemma~\ref{lem:noboundary} ensures that equation \eqref{eq:SteadyStateMF} with the boundary conditions \eqref{eq:LocalDensityLeft} and \eqref{eq:LocalDensityRight} is equivalent to a fixed-point equation, namely 
\begin{equation}
  \dfull_{n} = \varphi_n(\dfull_{1},\dots,\dfull_{N}) 
  \qquad n=1,\dots,N
  \, ,
  \label{eq:fixed_point_phi}
\end{equation}
where we have defined the functions
\begin{equation}
  \varphi_n(\dfull_{1},\dots,\dfull_{N}) \equiv 
  \frac
  {\hphantom{\oneminus{\dfull_{n+1}} +} \dfull_{n-1} 
	+ \omega_\mathrm{A} \hphantom{+ \omega_\mathrm{D}}}
  {\oneminus{\dfull_{n+1}} + \dfull_{n-1} 
	+ \omega_\mathrm{A} + \omega_\mathrm{D}} 
  \qquad n=1,\dots,N
  \, ,
\end{equation} 
with ${\dfull_{0} \equiv \alpha}$ and ${\dfull_{N+1} \equiv \oneminus{\beta}}$. 
Now, excluding the case ${\omega_\mathrm{A} + \omega_\mathrm{D} = 0}$ (i.e. ${\omega_\mathrm{A} = \omega_\mathrm{D} = 0}$), the functions ${\varphi_n: \mathbb{R}^{N} \to \mathbb{R}}$ are continuous over ${[0,1]}^{N}$, so that ${\varphi \equiv (\varphi_1,\dots,\varphi_N)}$ turns out to be a continuous mapping of the compact and convex set ${[0,1]}^{N}$ into itself.
Consequently, Brouwer's fixed-point theorem states that $\varphi$ has at least one fixed point ${(\dfull_1,\dots,\dfull_N) \in {[0,1]}^{N}}$.

\begin{lem}
  \label{lem:uniqueness}
  Let both the $N$-tuples ${(\dfull_{1},\dots,\dfull_{N}) \in {[0,1]}^{N}}$ and ${(\tilde{\dfull}_1,\dots,\tilde{\dfull}_N) \in {[0,1]}^{N}}$ be solutions of the stationary-state equation \eqref{eq:SteadyStateMF} with the boundary conditions \eqref{eq:LocalDensityLeft} and \eqref{eq:LocalDensityRight}. 
  Then $(\dfull_{1},\dots,\dfull_{N}) = (\tilde{\dfull}_1,\dots,\tilde{\dfull}_N)$. 
\end{lem}

\paragraph{Proof}  Summing \eqref{eq:SteadyStateMF} over $n$ with the boundary condition \eqref{eq:LocalDensityLeft}, we obtain
\begin{equation}
  \dfull_{n} \oneminus{\dfull_{n+1}} 
  = \alpha \oneminus{\dfull_{1}} 
  + \sum_{k=1}^{n} \left( 
  \omega_\mathrm{A} \oneminus{\dfull_{k}} - \omega_\mathrm{D} \dfull_{k} 
  \right)
  \qquad n=1,\dots,N
  \label{eq:rec}
\end{equation}
and the analogous equation
\begin{equation}
  \tilde{\dfull}_{n} \oneminus{\tilde{\dfull}_{n+1}\hphantom{}} 
  = \alpha \oneminus{\tilde{\dfull}_{1}\hphantom{}} 
  + \sum_{k=1}^{n} \left( 
  \omega_\mathrm{A} \oneminus{\tilde{\dfull}_{k}\hphantom{}} 
  - \omega_\mathrm{D} \tilde{\dfull}_{k} 
  \right)
  \qquad n=1,\dots,N
  \, .
  \label{eq:rec_tilde}
\end{equation}
We can divide both sides of \eqref{eq:rec} and \eqref{eq:rec_tilde} by $\dfull_{n}$ and $\tilde{\dfull}_{n}$, respectively, because Lemma~\ref{lem:noboundary} ensures that such quantities are nonzero. 
We argue that, given ${n\in\{1,\dots,N\}}$, if ${\tilde{\dfull}_{k} > \dfull_{k}}$ for all ${k=1,\dots,n}$ then ${\tilde{\dfull}_{n+1} > \dfull_{n+1}}$. 
We can then prove by induction that the condition ${\tilde{\dfull}_{1} > \dfull_{1}}$ implies ${\tilde{\dfull}_{n} > \dfull_{n}}$ for all ${n=2,\dots,N+1}$, and in particular ${\tilde{\dfull}_{N+1} > \dfull_{N+1}}$. 
This last inequality gives rise to a contradiction, because the boundary condition \eqref{eq:LocalDensityRight} imposes ${\tilde{\dfull}_{N+1} = \dfull_{N+1} = \oneminus{\beta}}$, so that the possibility ${\tilde{\dfull}_{1} > \dfull_{1}}$ must be excluded. 
An analogous argument (with reversed inequalities) allows us to exclude also the opposite case ${\tilde{\dfull}_{1} < \dfull_{1}}$, thus proving that ${\tilde{\dfull}_{1} = \dfull_{1}}$. 
Yet another analogous argument (with equalities replacing inequalities) shows that the condition ${\tilde{\dfull}_{1} = \dfull_{1}}$ implies ${\tilde{\dfull}_{n} = \dfull_{n}}$ for all $n$, which concludes the proof.

\begin{teo}
  \label{teo:EU}
  The stationary-state equation \eqref{eq:SteadyStateMF} with the boundary conditions \eqref{eq:LocalDensityLeft} and \eqref{eq:LocalDensityRight} admits a solution $(\dfull_{1},\dots,\dfull_{N}) \in {[0,1]}^{N}$, which cannot be a boundary point of ${[0,1]}^{N}$. 
  No other solution exists in the domain ${[0,1]}^{N}$. 
\end{teo}

\paragraph{Proof}  In the hypothesis ${\omega_\mathrm{A} + \omega_\mathrm{D} > 0}$, Lemmas \ref{lem:noboundary}, \ref{lem:existence} and~\ref{lem:uniqueness} immediately prove the theorem. 
Let us then consider the case ${\omega_\mathrm{A} + \omega_\mathrm{D} = 0}$ (i.e. ${\omega_\mathrm{A} = \omega_\mathrm{D} = 0}$), for which we still have to prove existence. 
For all positive integers $k$, let us define ${{\omega_{\mathrm{A}}}^{(k)} \equiv {\omega_{\mathrm{D}}}^{(k)} \equiv 1/k}$, so that for each $k$ we have ${{\omega_{\mathrm{A}}}^{(k)} + {\omega_{\mathrm{D}}}^{(k)} > 0}$, and thence a corresponding solution $({\dfull_{1}}^{(k)},\dots,{\dfull_{N}}^{(k)}) \in {[0,1]}^{N}$. 
The sequence of these solutions takes values in a compact set, which entails that it admits a convergent subsequence, in accordance to Bolzano-Weierstrass' theorem. 
The limit of such a subsequence is obviously a solution of \eqref{eq:SteadyStateMF} with \eqref{eq:LocalDensityLeft} and \eqref{eq:LocalDensityRight} in the case ${\omega_\mathrm{A} = \omega_\mathrm{D} = 0}$.

%% file: TASEP-LK_appendix-A.tex
\section{Bounds for the density and current profiles}
\label{app:Rho}

This appendix is devoted to the proof of Theorem~\ref{teo:delta} and Corollary~\ref{cor:current}. 
We shall always assume that our hypotheses \eqref{eq:intervallo_valori_Omega}, \eqref{eq:intervallo_valori_q0} and \eqref{eq:intervallo_valori_alpha} on the model parameters are verified, and they will often be taken into account without further reference.
Let us first prove some auxiliary results, which will be subsequently invoked several times.

\begin{lem}
  \label{lem:r_bounds}
  Let ${(\ddetr_{n})}_{n=0}^{N+1}$ be the sequence of detrended densities, defined as in section~\ref{sec:hd-phase}, with boundary values ${\ddetr_{0} = \alpha}$ and ${\ddetr_{N+1} = \dbulk_{0}}$.
  Then
  \begin{equation}
    \cases{
	  \alpha \geq \ddetr_{n} \geq \ddetr_{n+1} \geq \dbulk_{0}
	  & \textit{if} ${\alpha \geq \dbulk_{0}}$ \\
	  \alpha \leq \ddetr_{n} \leq \ddetr_{n+1} \leq \dbulk_{0}
	  & \textit{if} ${\alpha \leq \dbulk_{0}}$
    } 
    \qquad n=1,\dots,N-1
    \, ,
    \label{eq:bound_r}
  \end{equation} 
  \begin{equation}
    \cases{
	  \ddetr_{n-1}\oneminus{\ddetr_{n}} 
	  \geq \ddetr_{n}\oneminus{\ddetr_{n+1}} 
	  \geq \dbulk_{0}\oneminus{\dbulk_{0}}
	  & \textit{if} ${\alpha \geq \dbulk_{0}}$ \\
	  \ddetr_{n-1}\oneminus{\ddetr_{n}} 
	  \leq \ddetr_{n}\oneminus{\ddetr_{n+1}} 
	  \leq \dbulk_{0}\oneminus{\dbulk_{0}}
	  & \textit{if} ${\alpha \leq \dbulk_{0}}$
    }
    \qquad n = 1,\dots,N
    \, .
    \label{eq:bound_rr}
  \end{equation}
\end{lem}

\paragraph{Proof} From \eqref{eq:SteadyStateMF} with ${\omega_\mathrm{A} = \omega_\mathrm{D} = \omega}$, taking into account the definition \eqref{eq:detrended_densities_definition} of detrended densities, by simple algebra one arrives at the following equations
\begin{equation}
  \dfull_{n} \left( \ddetr_{n} - \ddetr_{n+1} \right)
  = \left( \ddetr_{n-1} - \ddetr_{n} \right) \oneminus{\dfull_{n}}
  \qquad n=1,\dots,N
  \, . 
  \label{eq:equazione_p-r}
\end{equation}
Now, relying on \eqref{eq:SteadyStateMF_bound}, i.e. Lemma~\ref{lem:noboundary}, we can deduce that the quantities ${\ddetr_{n} - \ddetr_{n+1}}$, for ${n=0,\dots,N}$, must all have the same sign or they must be all zero, so that the sequence ${(\ddetr_{n})}_{n=0}^{N+1}$ must be strictly monotonic or constant, respectively. 
We can thus write
\begin{equation}
  \cases{
	\ddetr_{0} \geq \ddetr_{n} \geq \ddetr_{n+1} \geq \ddetr_{N+1}
	& if ${\ddetr_{0} \geq \ddetr_{N+1}}$ \\
	\ddetr_{0} \leq \ddetr_{n} \leq \ddetr_{n+1} \leq \ddetr_{N+1}
	& if ${\ddetr_{0} \leq \ddetr_{N+1}}$
  } 
  \qquad n=1,\dots,N-1
  \, .
\end{equation} 
With the given boundary values, \eqref{eq:bound_r} follows immediately. 
Furthermore, from \eqref{eq:SteadyStateMF_detrended} we see that the condition ${\ddetr_{n-1} \geq \ddetr_{n+1}}$ implies ${\ddetr_{n-1}\oneminus{\ddetr_{n}} \geq \ddetr_{n} \oneminus{\ddetr_{n+1}}}$, for ${n=1,\dots,N}$, and the same with opposite inequalities. 
As a consequence we have 
\begin{equation}
  \cases{
	\ddetr_{n-1}\oneminus{\ddetr_{n}} 
	\geq \ddetr_{n} \oneminus{\ddetr_{n+1}} 
	\geq \ddetr_{N} \oneminus{\ddetr_{N+1}}
	& $\!\!\!\!\!\!\!$ if ${\ddetr_{0} \geq \ddetr_{N+1}}$ \\
	\ddetr_{n-1}\oneminus{\ddetr_{n}} 
	\leq \ddetr_{n} \oneminus{\ddetr_{n+1}} 
	\leq \ddetr_{N} \oneminus{\ddetr_{N+1}}
	& $\!\!\!\!\!\!\!$ if ${\ddetr_{0} \leq \ddetr_{N+1}}$
  } 
  \quad n=1,\dots,N-1
  \, .
\end{equation} 
With the given boundary value ${\ddetr_{N+1} = \dbulk_{0}}$, by means of \eqref{eq:bound_r} and ${\oneminus{\dbulk_{0}} > 0}$, we finally obtain \eqref{eq:bound_rr}. 

\begin{lem}
	\label{lem:qq}
	Let ${(\dbulk_{n})}_{n=0}^{N+1}$ be the sequence of bulk densities, defined as in section~\ref{sec:hd-phase}. 
	Then, the following inequalities hold
	\begin{equation}
	\dbulk_{0}\oneminus{\dbulk_{0}} \geq 
	\dbulk_{n-1}\oneminus{\dbulk_{n}} \geq 
	\dbulk_{n}\oneminus{\dbulk_{n+1}} \geq 
	\dbulk_{0}\oneminus{\dbulk_{0}} - C \omega (n+1)
	\qquad n=1,\dots,N
	\, .
	\end{equation}
\end{lem}

\paragraph{Proof} Using the bulk-density expression \eqref{eq:bulk_densities_expression}, i.e. ${\dbulk_{n} = \dbulk_{0} + \omega{n}}$, we see that
\begin{equation}
  \dbulk_{n} \oneminus{\dbulk_{n+1}} 
  = \dbulk_{0} \oneminus{\dbulk_{0}} 
  - \omega \dbulk_{0} 
  - \omega n \left[\dbulk_{0}-\oneminus{\dbulk_{0}} + \omega (n+1)\right]
  \qquad n=0,\dots,N
  \, , \quad
  \label{eq:qq}
\end{equation}
where we notice that ${\dbulk_{0} - \oneminus{\dbulk_{0}} > 0}$, since by hypothesis ${\dbulk_{0} > 1/2}$. Consequently, ${\dbulk_{n} \oneminus{\dbulk_{n+1}}}$ is either constant (for ${\Omega = 0}$) or decreasing in $n$ (for ${\Omega > 0}$), and the upper-bound is evident. The lower-bound easily follows observing that ${\omega (n+1) \leq \Omega}$. 

\subsection{Proof of Theorem~\ref{teo:delta}}

Let us define the sequences
\begin{eqnarray}
  \varrho_{n} & \equiv
  \cases{
  	\ddetr_{n} - \dbulk_{0} & if ${\alpha \geq \dbulk_{0}}$ \\
  	\dbulk_{0} - \ddetr_{n} & if ${\alpha \leq \dbulk_{0}}$
  }
  & \qquad n = 0,\dots,N+1
  \, , 
  \label{eq:definizione_rho} 
  \\
  \varepsilon_{n} & \equiv
  \cases{
	\ddetr_{n}\oneminus{\ddetr_{n+1}} - \dbulk_{0}\oneminus{\dbulk_{0}} 
	& if ${\alpha \geq \dbulk_{0}}$ \\
	\dbulk_{0}\oneminus{\dbulk_{0}} - \ddetr_{n}\oneminus{\ddetr_{n+1}} 
	& if ${\alpha \leq \dbulk_{0}}$
  }
  & \qquad n = 0,\dots,N
  \, , 
  \label{eq:definizione_epsilon} 
  \\
  \delta_{n} & \equiv
  \cases{
    \ddlim_{n} - \ddetr_{n} 
	& if ${\alpha \geq \dbulk_{0}}$ \\
    \ddetr_{n} - \ddlim_{n} 
	& if ${\alpha \leq \dbulk_{0}}$
  }
  & \qquad n = 0,\dots,N+1
  \, . 
  \label{eq:definizione_delta} 
\end{eqnarray}
Note that the above definitions remain consistent even in the case ${\alpha = \dbulk_{0}}$, because, according to \eqref{eq:definizione_s} in Lemma \ref{lem:eta} and \eqref{eq:bound_r} in Lemma~\ref{lem:r_bounds}, in that case we have ${\ddetr_{n} = \ddlim_{n} = \dbulk_{0}}$ for all $n$, and all three sequences turn out to be constantly equal to zero. 
Now, from \eqref{eq:definizione_rho} and \eqref{eq:bound_r} we get 
\begin{equation}
  \varrho_{n} \geq 0 \qquad n = 0,\dots,N+1
  \, , 
  \label{eq:lowerbound_rho}
\end{equation}
which also allows us to write
\begin{equation}
  \varrho_{n} =
  \left|
    \ddetr_{n} - \dbulk_{0}
  \right|
  \qquad n = 0,\dots,N+1
  \, .
  \label{eq:rho_modulo}
\end{equation}
Moreover, from \eqref{eq:definizione_epsilon} and \eqref{eq:bound_rr} we get
\begin{equation}
  \varepsilon_{n} \geq 0 
  \qquad n = 0,\dots,N
  \, ,
  \label{eq:lowerbound_epsilon}
\end{equation}
which also allows us to write
\begin{equation}
  \varepsilon_{n} =
  \left|
	\ddetr_{n}\oneminus{\ddetr_{n+1}} - \dbulk_{0}\oneminus{\dbulk_{0}} 
  \right|
  \qquad n = 0,\dots,N
  \, .
  \label{eq:epsilon_modulo}
\end{equation}
Furthermore, using \eqref{eq:definizione_epsilon}, \eqref{eq:definizione_delta} and \eqref{eq:proposizione_eta} (Lemma~\ref{lem:eta}), we can write
\begin{equation}
  \varepsilon_{n} = \ddetr_{n} \delta_{n+1} - \oneminus{\ddlim_{n+1}} \delta_{n}
  \qquad n = 0,\dots,N
  \, . 
  \label{eq:relazione_epsilon-delta}
\end{equation}
By means of \eqref{eq:bound_r} in Lemma~\ref{lem:r_bounds}, along with ${\dbulk_{0} > 0}$ and ${\alpha > 0}$, we see that 
$\ddetr_{n}$ is a positive quantity for all $n$, so that from \eqref{eq:lowerbound_epsilon} and~\eqref{eq:relazione_epsilon-delta} we can write
\begin{equation}
  \delta_{n+1} \geq \frac{\oneminus{\ddlim_{n+1}}}{\ddetr_{n}} \delta_{n}
  \qquad n = 0,\dots,N
  \, ,
\end{equation}
where Lemma \ref{lem:eta} ensures that the term ${\oneminus{\ddlim_{n+1}}/\ddetr_{n}}$ is positive as well for all $n$.
Moreover, by Lemma \ref{lem:eta} we see that ${\ddlim_{0} = \alpha}$, so with ${\ddetr_{0} = \alpha}$ we have ${\delta_{0} = 0}$. 
Then, applying recursively the above inequality, we arrive at
\begin{equation}
  \delta_{n} \geq 0 \qquad n=0,\dots,N+1
  \, ,
  \label{eq:lowerbound_delta}
\end{equation}
which also allows us to write
\begin{equation}
  \delta_{n} = \left| 
	\ddlim_{n} - \ddetr_{n} 
  \right|
  \qquad n = 0,\dots,N+1
  \, . 
  \label{eq:delta_modulo}
\end{equation}

So far, we have proved that $\varrho_{n}$, $\varepsilon_{n}$ and $\delta_{n}$ are all non-negative quantities. 
We now prove upper-bounds for the same quantities. 
Before entering the details, let us give a sketch of the main steps needed for this proof. 
First, making use of Lemma~\ref{lem:eta} and of the lower-bounds \eqref{eq:lowerbound_rho} and~\eqref{eq:lowerbound_delta}, we prove the upper-bounds for $\varrho_{n}$ and $\varepsilon_{n}$. 
Subsequently, making use of the upper-bound for $\varepsilon_{n}$ together with \eqref{eq:relazione_epsilon-delta}, we also get the upper-bound for $\delta_{n}$, i.e. the thesis.

\paragraph{Upper-bound for $\varrho_{n}$}

From Lemma~\ref{lem:eta} we can easily deduce the following bounds
\begin{equation}
  \cases{
	\ddlim_{n} \leq \dbulk_{0} + C \gamma^{n} 
	& if ${\alpha \geq \dbulk_{0}}$ \\
	\ddlim_{n} \geq \dbulk_{0} - C \gamma^{n} 
	& if ${\alpha \leq \dbulk_{0}}$
  }
  \qquad n = 0,1,2,\dots
  \, .
  \label{eq:bound-s}
\end{equation}
Moreover, thanks to \eqref{eq:lowerbound_delta} and \eqref{eq:definizione_delta}, we can write
\begin{equation}
  \cases{
    \ddetr_{n} \leq \ddlim_{n} 
    & if ${\alpha \geq \dbulk_{0}}$ \\
    \ddetr_{n} \geq \ddlim_{n} 
    & if ${\alpha \leq \dbulk_{0}}$
  }
  \qquad n = 0,\dots,N+1
  \, ,
  \label{eq:disuguaglianza_r-s}
\end{equation}
and therefore, with definition~\eqref{eq:definizione_rho}, 
\begin{equation}
  \varrho_{n} \leq C \gamma^{n}
  \qquad n = 0,\dots,N+1
  \, .
  \label{eq:upperbound_rho}
\end{equation}
Taking into account \eqref{eq:rho_modulo}, this is equivalent to 
\begin{equation}
  \left| \ddetr_{n} - \dbulk_{0} \right| 
  \leq C \gamma^{n}
  \qquad n = 0,\dots,N+1 
  \, . 
  \label{eq:upperbound_rho_modulo} 
\end{equation}

\paragraph{Upper-bound for $\varepsilon_{n}$}

Let us observe that one can write, in full generality, 
\begin{equation}
  \varepsilon_{n}
  = \sum_{k=n+1}^{N} \left( \varepsilon_{k-1} - \varepsilon_{k} \right)
  + \varepsilon_{N}
  \qquad n = 0,\dots,N-1
  \, .
  \label{eq:somma_telescopica}
\end{equation}
Taking into account, in order, \eqref{eq:SteadyStateMF_detrended} with \eqref{eq:definizione_rho} and~\eqref{eq:definizione_epsilon}, \eqref{eq:lowerbound_rho}, and finally \eqref{eq:upperbound_rho}, we have
\begin{eqnarray}
  \varepsilon_{n-1} - \varepsilon_{n}
  = \omega n \left( \varrho_{n-1} - \varrho_{n+1} \right)
  \leq \omega n \varrho_{n-1} 
  \leq C \omega n \gamma^{n-1}
  \qquad n = 1,\dots,N
  \, .
  \nonumber \\
  \label{eq:somma_telescopica_elemento}
\end{eqnarray}
Moreover, using \eqref{eq:definizione_rho} and~\eqref{eq:definizione_epsilon} with ${\ddetr_{N+1} = \dbulk_{0}}$, and then \eqref{eq:upperbound_rho} with ${\oneminus{\dbulk_{0}} > 0}$, we have
\begin{equation}
  \varepsilon_{N} = \varrho_{N} \oneminus{\dbulk_{0}} \leq C \gamma^{N}
  \, . 
  \label{eq:upperbound_epsilon_N}
\end{equation}
We now need to distinguish the pure TASEP from the TASEP-LK. In the former case we have ${\omega = 0}$, so that \eqref{eq:somma_telescopica_elemento} reads ${\varepsilon_{n-1} - \varepsilon_{n} = 0}$. Then from \eqref{eq:somma_telescopica} and \eqref{eq:upperbound_epsilon_N} we get 
\begin{equation}
  \varepsilon_{n}
  \leq C \gamma^{N} 
  \qquad n = 0,\dots,N
  \, .
  \label{eq:upperbound_epsilon_omegaeq0}
\end{equation}
Otherwise, in the TASEP-LK case, we have ${\omega = \Omega/(N+1)}$ with ${\Omega > 0}$, so that \eqref{eq:somma_telescopica_elemento} reads ${\varepsilon_{n-1} - \varepsilon_{n} \leq C n \gamma^{n-1}/(N+1)}$. Then from \eqref{eq:somma_telescopica} and \eqref{eq:upperbound_epsilon_N} we get 
\begin{equation}
  \varepsilon_{n}
  \leq \frac{C}{N+1} \sum_{k=n+1}^{N+1} k \gamma^{k-1} 
  \qquad n = 0,\dots,N
  \, .
\end{equation}
Recalling \eqref{eq:intervallo_valori_gamma}, the sum can be manipulated as follows
\begin{equation}
  \sum_{k=n+1}^{N+1} k \gamma^{k-1} 
  < \sum_{k=n+1}^{\infty} k \gamma^{k-1}
  = \frac{(n+1)\gamma^{n} - n\gamma^{n+1}}{(1-\gamma)^{2}}
  \, .
\end{equation}
We then have 
\begin{equation}
  \varepsilon_{n}
  \leq C \frac{(n+1)\gamma^{n}}{N}
  \qquad n = 0,\dots,N
  \, ,
  \label{eq:upperbound_epsilon_omegagt0}
\end{equation}
where we have legitimately replaced ${N+1}$ with $N$ in the denominator. 
We have thus arrived at the announced upper-bounds for $\varepsilon_{n}$, namely \eqref{eq:upperbound_epsilon_omegaeq0} for ${\Omega = 0}$ and \eqref{eq:upperbound_epsilon_omegagt0} for ${\Omega > 0}$. 
Taking into account \eqref{eq:epsilon_modulo}, they are equivalent to 
\begin{equation}
  \left| \ddetr_{n}\oneminus{\ddetr_{n+1}} - \dbulk_{0}\oneminus{\dbulk_{0}} \right|
  \leq \cases{ 
  	C \gamma^{N}                 & if ${\Omega = 0}$ \\
  	C \frac{(n+1) \gamma^{n}}{N} & if ${\Omega > 0}$
  } \qquad n = 0,\dots,N
  \, . 
  \label{eq:upperbound_epsilon_modulo} 
\end{equation}

\paragraph{Upper-bound for $\delta_{n}$}

Let us replace $\varepsilon_{n}$ with the expression given by \eqref{eq:relazione_epsilon-delta}. 
Taking into account that, by virtue of Lemma~\ref{lem:r_bounds} with ${\dbulk_{0} > 0}$ and ${\alpha > 0}$, the quantities $\ddetr_{n}$ are bounded from below by a positive constant, we can write, for ${\Omega = 0}$, 
\begin{equation}
  \delta_{n+1} \leq \frac{\oneminus{\ddlim_{n+1}}}{\ddetr_{n}} \delta_{n}
  + C \gamma^{N} 
  \qquad n = 0,\dots,N
  \, ,
  \label{eq:disuguaglianza_delta_omegaeq0}
\end{equation}
and, for ${\Omega > 0}$,
\begin{equation}
  \delta_{n+1} \leq \frac{\oneminus{\ddlim_{n+1}}}{\ddetr_{n}} \delta_{n}
  + C \frac{(n+1) \gamma^{n}}{N} 
  \qquad n = 0,\dots,N
  \, ,
  \label{eq:disuguaglianza_delta_omegagt0}
\end{equation}
where, by Lemmas \ref{lem:eta} and \ref{lem:r_bounds}, we know that the term ${\oneminus{\ddlim_{n+1}}/\ddetr_{n}}$ is always positive and bounded by a constant. 
Let us first consider the case ${\Omega = 0}$. 
Recalling that ${\delta_{0} = 0}$ and applying recursively \eqref{eq:disuguaglianza_delta_omegaeq0}, we easily obtain 
\begin{equation}
  \delta_{n} \leq C \gamma^{N} 
  \qquad n = 0,\dots,N+1
  \, .
\end{equation}
Taking into account \eqref{eq:delta_modulo}, this is equivalent to the thesis of Theorem~\ref{teo:delta}, for the case ${\Omega = 0}$.
The case ${\Omega > 0}$ is slightly more complicated. Applying recursively \eqref{eq:disuguaglianza_delta_omegagt0}, we first obtain
\begin{equation}
  \delta_{n} \leq \frac{C}{N} \sum_{k=0}^{n} k \gamma^{k-1} \prod_{l=k}^{n-1} \frac{\oneminus{\ddlim_{l+1}}}{\ddetr_{l}}
  \qquad n = 0,\dots,N+1
  \, . 
  \label{eq:upperbound_delta_intermedio}
\end{equation}
From now on we need to distinguish whether ${\alpha \geq \dbulk_{0}}$ or ${\alpha < \dbulk_{0}}$. In the former case, from \eqref{eq:disuguaglianza_r-s} and \eqref{eq:bound_r}, along with ${\gamma = \oneminus{\dbulk_{0}}/\dbulk_{0}}$, we easily get
\begin{equation}
  \frac{\oneminus{\ddlim_{n+1}}}{\ddetr_{n}} \leq \gamma
  \qquad n = 0,\dots,N
  \, , 
  \label{eq:disuguaglianza_fattore}
\end{equation}
and therefore
\begin{equation}
  \delta_{n} \leq \frac{C}{N} \left( \sum_{k=0}^{n} k \right) \gamma^{n-1}
  \leq C \frac{n^2 \gamma^{n}}{N} 
  \qquad n = 0,\dots,N+1
  \, .
  \label{eq:upperbound_delta}
\end{equation}
Taking into account \eqref{eq:delta_modulo}, this is equivalent to the thesis of Theorem~\ref{teo:delta}, for the case ${\Omega > 0}$. 
At this point, to complete the proof we only need to prove that \eqref{eq:upperbound_delta} holds even in the complementary hypothesis ${\alpha < \dbulk_{0}}$. 
In such a case, \eqref{eq:disuguaglianza_r-s} and \eqref{eq:bound_r} imply the opposite of \eqref{eq:disuguaglianza_fattore}, so that the above argument cannot be applied. 
Nevertheless, we can observe that, because of \eqref{eq:bound-s} and~\eqref{eq:disuguaglianza_r-s}, both $\ddlim_{n}$ and $\ddetr_{n}$ approach $\dbulk_{0}$ exponentially, upon increasing $n$. 
As a consequence, the term
${\oneminus{\ddlim_{n+1}}/\ddetr_{n}}$ approaches $\gamma$ exponentially as well, which entails that in \eqref{eq:upperbound_delta_intermedio} it should be possible to write
\begin{equation}
  \prod_{l=k}^{n-1} \frac{\oneminus{\ddlim_{l+1}}}{\ddetr_{l}}
  \leq C \gamma^{n-k}
  \, ,
\end{equation}
so that we can still obtain~\eqref{eq:upperbound_delta}.

\subsection{Proof of Corollary~\ref{cor:current}}

Let us recall the current-density relationship ${{J}_{n} = \dfull_{n} \oneminus{\dfull_{n+1}}}$, the definition of detrended densities \eqref{eq:detrended_densities_definition}, i.e. ${\dfull_{n} = \ddetr_{n} + \omega{n}}$, and the bulk solution \eqref{eq:bulk_densities_expression}, i.e. ${\dbulk_{n} = \dbulk_{0} + \omega{n}}$. Simple algebra yields
\begin{equation}
  {J}_{n} - \dbulk_{n} \oneminus{\dbulk_{n+1}} 
  = ( \ddetr_{n} \oneminus{\ddetr_{n+1}} - \dbulk_{0} \oneminus{\dbulk_{0}} ) 
  - \omega n ( \ddetr_{n+1} - \dbulk_{0} )
  - \omega (n+1) ( \ddetr_{n} - \dbulk_{0} )
  \, , \quad
\end{equation}
and thence, using the triangular inequality, 
\begin{equation}
  | {J}_{n} - \dbulk_{n} \oneminus{\dbulk_{n+1}} | 
  \leq
  | \ddetr_{n} \oneminus{\ddetr_{n+1}} - \dbulk_{0} \oneminus{\dbulk_{0}} | 
  + \omega n | \ddetr_{n+1} - \dbulk_{0} |
  + \omega (n+1) | \ddetr_{n} - \dbulk_{0} |
  \, . \quad
\end{equation}
Both the above statements hold for ${n=0,\dots,N}$.
Then, using \eqref{eq:upperbound_rho_modulo} and \eqref{eq:upperbound_epsilon_modulo} from inside the proof of Theorem~\ref{teo:delta}, we easily prove \eqref{eq:bound_current}.
In order to prove also \eqref{eq:bound_current-max}, let us first note that, using again the triangular inequality, we can write
\begin{eqnarray}
  \left| 
    \max {({J}_{n})}_{n=0}^{N} 
    - \dbulk_{0}\oneminus{\dbulk_{0}}
  \right| \leq & \left| 
    \max {({J}_{n})}_{n=0}^{N} 
    - \max {(\dbulk_{n}\oneminus{\dbulk_{n+1}})}_{n=0}^{N}
  \right| + \nonumber \\ & + \left| 
    \max {(\dbulk_{n}\oneminus{\dbulk_{n+1}})}_{n=0}^{N} 
    - \dbulk_{0}\oneminus{\dbulk_{0}}
  \right| 
  \, .
\end{eqnarray}
Now, for two generic sequences ${({x}_{n})}_{n=0}^{N}$ and ${({y}_{n})}_{n=0}^{N}$, the following inequality holds
\begin{equation}
  \left| 
    \max {({x}_{n})}_{n=0}^{N}
    - \max {({y}_{n})}_{n=0}^{N}
  \right| 
  \leq \max {(|{x}_{n} - {y}_{n}|)}_{n=0}^{N}
  \, ,
\end{equation}
whereas Lemma~\ref{lem:qq} obviously entails
\begin{equation}
  \max {(\dbulk_{n}\oneminus{\dbulk_{n+1}})}_{n=0}^{N} 
  = \dbulk_{0}\oneminus{\dbulk_{1}}
  \, .
\end{equation}
As a consequence we can write
\begin{equation}
  \left| 
    \max {({J}_{n})}_{n=0}^{N} 
    - \dbulk_{0}\oneminus{\dbulk_{0}}
  \right| 
  \leq \max {(|
	{J}_{n} - \dbulk_{n}\oneminus{\dbulk_{n+1}}
  |)}_{n=0}^{N}
  + \omega \dbulk_{0} 
  \, . 
\end{equation}
The proof of \eqref{eq:bound_current-max} follows easily from \eqref{eq:bound_current}.

%% file: TASEP-LK_appendix-B.tex
\section{Bounds for the slowest relaxation rate}
\label{app:Lambda}

\subsection{Proof of Corollary~\ref{cor:simplified-matrix}}

The proof relies on the \emph{Courant minimax principle}, by which we can state both ${\lambda = \min{\{({u},{A}{u})\}}_{{u} \in \mathbb{R}^{N} : \|{u}\|=1}}$ and ${\mu = \min{\{({u},{B}{u})\}}_{{u} \in \mathbb{R}^{N} : \|{u}\|=1}}$, and therefore
\begin{equation}
  | \lambda - \mu | \leq 
  \max{\{|({u},{A}{u}) - ({u},{B}{u})|\}}_{{u} \in \mathbb{R}^{N} : \|{u}\|=1}
  \, .
\end{equation}
Thus, Corollary~\ref{cor:simplified-matrix} follows if we prove the bound for ${|({u},{A}{u}) - ({u},{B}{u})|}$, where
\begin{eqnarray}
  ({u},{A}{u}) - ({u},{B}{u}) 
  = & \sum_{n=1}^{N} \left( 
    \ddlim_{n+1} - \ddetr_{n+1} - \ddlim_{n-1} + \ddetr_{n-1}
  \right) {{u}_{n}}^{2} 
  + \nonumber \\ &
  - 2 \sum_{n=1}^{N-1} \left( 
    \sqrt{\dfull_{n} \oneminus{\dfull_{n+1}}} - \sqrt{\dbulk_{n} \oneminus{\dbulk_{n+1}}} 
  \right) {u}_{n} {u}_{n+1}  
  \, .
\end{eqnarray}
To this aim, we first note that, thanks to the triangular inequality, we can write
\begin{eqnarray}
  \left| ({u},{A}{u}) - ({u},{B}{u}) \right|
  \leq & \sum_{n=1}^{N} \left( 
  \left| \ddlim_{n+1} - \ddetr_{n+1} \right| +
  \left| \ddlim_{n-1} - \ddetr_{n-1} \right|
  \right) {{u}_{n}}^{2} 
  + \nonumber \\ &
  + 2 \sum_{n=1}^{N-1} \left| 
    \sqrt{\dfull_{n} \oneminus{\dfull_{n+1}}} - \sqrt{\dbulk_{n} \oneminus{\dbulk_{n+1}}} 
  \right| \left|{u}_{n}\right| \left|{u}_{n+1}\right| 
  \, ,
  \label{eq:upperbound_uAu-uBu}
\end{eqnarray}
where from Theorem~\ref{teo:delta} we have, for all $n$, 
\begin{equation}
  \left| \ddlim_{n+1} - \ddetr_{n+1} \right| +
  \left| \ddlim_{n-1} - \ddetr_{n-1} \right|
  \leq \cases{ 
    C \gamma^{N} & if ${\Omega = 0}$ \\
    C {N}^{-1}   & if ${\Omega > 0}$ 
  } 
  \, .
  \label{eq:upperbound_a-b}
\end{equation}
Moreover, from Corollary~\ref{cor:current} we obtain, for all $n$, 
\begin{equation}
  \left| 
    \dfull_{n} \oneminus{\dfull_{n+1}} - \dbulk_{n} \oneminus{\dbulk_{n+1}} 
  \right| 
  \leq \cases{ 
  	C \gamma^{N} & if ${\Omega = 0}$ \\
  	C {N}^{-1}   & if ${\Omega > 0}$ 
  } 
  \, ,
  \label{eq:upperbound_pp-qq}
\end{equation}
from which the analogous bound for ${\left| \sqrt{\dfull_{n} \oneminus{\dfull_{n+1}}} - \sqrt{\dbulk_{n} \oneminus{\dbulk_{n+1}}} \right|}$ can be easily proved. 
At this point, we pick any vector $u$ of unit norm. This means that
${\sum_{n=1}^{N} {{u}_{n}}^{2} = 1}$, giving in particular
${\sum_{n=1}^{N-1} \left|{u}_{n}\right| \left|{u}_{n+1}\right| \le 1}$, thanks to the \emph{Cauchy-Schwarz inequality}. Then, invoking \eqref{eq:upperbound_uAu-uBu}, \eqref{eq:upperbound_a-b} and \eqref{eq:upperbound_pp-qq}, the proof is easily concluded.

\subsection{Proof of Lemma \ref{lem:Gershgorin-type_bound}}
\label{app:Gershgorin-type_bound}

Let us denote by $v$ an eigenvector of $B$ corresponding to the smallest eigenvalue $\mu$, so that ${B v = \mu v}$. 
The vector $v$ can be chosen in such a way that ${v_{n} = w_{n}}$ for some ${n \in \{1,\dots,N\}}$ and ${|v_{k}| \leq w_{k}}$ for the other ${k \neq n}$ (even ${k = 0}$ and ${k = N+1}$, by setting ${v_{0} \equiv v_{N+1} \equiv 0}$). 
In practice, one can take $n$ to be an index where the maximum of ${\{|v_{1}|/w_1,\dots,|v_{N}|/w_{N}\}}$ is attained, and then normalize $v$ by ${v_{n}/w_{n}}$. 
From this argument, we have in particular that ${v_{n} \neq 0}$, so that we can write ${\mu = {Bv_{n}}/{v_{n}}}$, that is
\begin{equation}
  \mu = 1 - \sqrt{\dbulk_{0} \oneminus{\dbulk_{0}}} 
  \left( \sigma_{n+1} - \sigma_{n-1} \right)
  - \sqrt{\dbulk_{n} \oneminus{\dbulk_{n+1}}} \, \frac{{v}_{n+1}}{v_{n}} 
  - \sqrt{\dbulk_{n-1} \oneminus{\dbulk_{n}}} \, \frac{{v}_{n-1}}{v_{n}} 
  \, , 
\end{equation}
where we have also used \eqref{eq:definizione_sigma_vecchia}. 
Then, using ${v_{n} = w_{n} > 0}$ and ${v_{n \pm 1} \leq w_{n \pm 1}}$, we easily get the following bound
\begin{equation}
  \mu \geq 1 - \sqrt{\dbulk_{0} \oneminus{\dbulk_{0}}} 
  \left( \sigma_{n+1} - \sigma_{n-1} \right) 
  - \sqrt{\dbulk_{n} \oneminus{\dbulk_{n+1}}} \, \frac{{w}_{n+1}}{{w}_{n}} 
  - \sqrt{\dbulk_{n-1} \oneminus{\dbulk_{n}}} \, \frac{{w}_{n-1}}{{w}_{n}} 
\, , 
\end{equation}
which, observing that from Lemma~\ref{lem:qq} we have ${\dbulk_{n} \oneminus{\dbulk_{n+1}} \leq \dbulk_{n-1} \oneminus{\dbulk_{n}} \leq \dbulk_{0} \oneminus{\dbulk_{0}}}$, simplifies to
\begin{equation}
  \mu \geq 1 - \sqrt{\dbulk_{0} \oneminus{\dbulk_{0}}} 
  \left(
    \sigma_{n+1} - \sigma_{n-1} + \frac{{w}_{n+1} + {w}_{n-1}}{{w}_{n}} 
  \right)
  \, .
\end{equation}
At this point, we see that in general one cannot foresee the proper index ${n}$, so that we are forced to choose the worst case, and this concludes the proof.

\subsection{Proof of Lemma~\ref{lem:upperbound_mu}}
\label{app:upperbound_mu}

With the aim of using the Courant-type bound (Lemma~\ref{lem:Courant-type_bound}), let us define ${{u}_{1},\dots,{u}_{N}}$ as
\begin{equation}
  {u}_{n} \equiv \cases{
    \sqrt{\frac{2}{M}} \, \sin \frac{\pi n}{M} & if ${n \leq M}$ \\
	0 & if ${n \geq M}$
  }
  \, ,
  \label{eq:trial-vector}
\end{equation}
where ${M \geq 2}$ is an integer. 
We hypotesize ${M \leq N+1}$, so that the above definition verifies ${\sum_{n=1}^{N} {{u}_{n}}^{2} = 1}$, as required by Lemma~\ref{lem:Courant-type_bound}. 
Observing that in \eqref{eq:trial-vector} we have ${{u}_{n} = 0}$ for ${n \geq M}$, from \eqref{eq:Courant-type_bound} we obtain
\begin{equation}
  \mu \leq 1 
  + \sqrt{\dbulk_{0} \oneminus{\dbulk_{0}}} 
  \, \sum_{n=1}^{M-1} \left( \sigma_{n-1} - \sigma_{n+1} \right) {{u}_{n}}^{2}
  - 2 \sum_{n=1}^{M-2} \sqrt{\dbulk_{n} \oneminus{\dbulk_{n+1}}} \, {u}_{n} {u}_{n+1}
  \, .
  \label{case10}
\end{equation}
From Lemma~\ref{lem:eta} we can easily deduce the bound ${\sigma_{n-1} - \sigma_{n+1} \leq C \gamma^{n}}$, whereas ${\sin x \leq x}$ for all real non-negative $x$, so that we can write
\begin{equation}
  \sum_{n=1}^{M-1} \left( \sigma_{n-1} - \sigma_{n+1} \right) {{u}_{n}}^2
  \leq \frac{C}{M^3}
  \, .
  \label{case12}
\end{equation}
Furthermore, by Lemma~\ref{lem:qq} we see that ${\dbulk_{n} \oneminus{\dbulk_{n+1}} \geq \dbulk_{M-2} \oneminus{\dbulk_{M-1}}}$ for all ${n = 1,\dots,M-2}$. 
Thence, observing also that ${\sum_{n=1}^{M-2} {u}_{n} {u}_{n+1} = \cos{(\pi/M)}}$, we get 
\begin{equation}
  \sum_{n=1}^{M-2} \sqrt{\dbulk_{n} \oneminus{\dbulk_{n+1}}} \, {u}_{n} {u}_{n+1}
  \geq \sqrt{\dbulk_{M-2} \oneminus{\dbulk_{M-1}}} \, \cos \frac{\pi}{M}
  \, .
  \label{case11}
\end{equation}
This way, combining \eqref{case11} and \eqref{case12} with \eqref{case10}, we reach the inequality
\begin{equation}
  \mu \leq 1 + \frac{C}{M^3} 
  - 2 \sqrt{\dbulk_{M-2} \oneminus{\dbulk_{M-1}}} \, \cos \frac{\pi}{M}
  \, .
  \label{eq:upperbound_mu_intermedio}
\end{equation}
Using again Lemma~\ref{lem:qq}, we easily show that ${\sqrt{\dbulk_{M-2} \oneminus{\dbulk_{M-1}}} \geq \sqrt{\dbulk_{0} \oneminus{\dbulk_{0}}} - C \Omega {M}/{N}}$, whereas ${\cos{(\pi/M)} \geq 1 - \frac{1}{2} {(\pi/{M})}^{2}}$, so that from \eqref{eq:upperbound_mu_intermedio} we obtain
\begin{equation}
  \mu \leq 1 - 2 \sqrt{\dbulk_{0} \oneminus{\dbulk_{0}}} 
  + \pi^{2} \sqrt{\dbulk_{0} \oneminus{\dbulk_{0}}} \, {M}^{-2}
  + C \Omega {M} {N}^{-1} 
  + C {M}^{-3}
  \, .
\end{equation}
Let us finally distinguish the pure TASEP from the TASEP-LK. In the former case we have ${\Omega = 0}$ and the term in ${{M}{N}^{-1}}$ disappears. Then the most restrictive bound can be attained choosing $M$ as large as possible, i.e. ${M = N+1}$, which allows us to write
\begin{equation}
  \mu \leq 1 - 2 \sqrt{\dbulk_{0} \oneminus{\dbulk_{0}}} 
  + \pi^{2} \sqrt{\dbulk_{0} \oneminus{\dbulk_{0}}} \, {N}^{-2}
  + C {N}^{-3}
  \, .
\end{equation}
The term in ${{N}^{-3}}$ may be dropped, provided the prefactor of ${{N}^{-2}}$ is replaced by a \emph{suitable constant} $C$. 
Otherwise, if ${\Omega > 0}$, the terms in ${M}^{-2}$ and ${M} {N}^{-1}$ are both present, and one can easily argue that the most restrictive bound is attained when they have the same asymptotic order, that is choosing ${M}$ of the order of ${{N}^{1/3}}$, which leads to  
\begin{equation}
  \mu \leq 1 - 2 \sqrt{\dbulk_{0} \oneminus{\dbulk_{0}}} 
  + C {N}^{-2/3} 
  + C {N}^{-1}
  \, .
\end{equation}
The term in ${{N}^{-1}}$ may obviously be dropped, and this concludes the proof.

\subsection{Proof of Lemma~\ref{lem:successione_v}}
\label{app:successione_v}

\input{TASEP-LK_appendix-B4.tex}

\subsection{Proof of Lemma~\ref{lem:upperbound_mu_alfa_piccoli}}
\label{app:upperbound_mu_alfa_piccoli}

Let us consider the Courant-type bound (Lemma~\ref{lem:Courant-type_bound}), and let us define 
\begin{equation}
  {u}_{n} \equiv \frac{{v}_{n}({x}_{*})}{\sqrt{\sum_{k=1}^{N} {{v}_{k}({x}_{*})}^{2}}}
  \qquad n=0,\dots,N+1
  \, ,
  \label{eq:definizione_u}
\end{equation}
${v}_{n}({x}_{*})$ being defined according to Lemma~\ref{lem:successione_v}. 
Note that \eqref{eq:definizione_u} entails ${\sum_{n=1}^{N} {{u}_{n}}^{2} = 1}$, as required by Lemma~\ref{lem:Courant-type_bound}, and also, due to Lemma~\ref{lem:successione_v}, ${{u}_{n} > 0}$ for all ${n > 0}$ and ${{u}_{0} = 0}$.
Moreover, taking into account Lemma~\ref{lem:qq}, we easily obtain the bound
\begin{equation}
  \sqrt{\dbulk_{n}\oneminus{\dbulk_{n+1}}} \geq 
  \sqrt{\dbulk_{0}\oneminus{\dbulk_{0}}} - C \omega (n+1)
  \qquad n=0,\dots,N
  \, .
\end{equation}
As a consequence, from \eqref{eq:Courant-type_bound} we can write
\begin{equation}
\fl
  \mu \leq 1 
  - \sqrt{\dbulk_{0} \oneminus{\dbulk_{0}}} \left[
    \sum_{n=1}^{N} \left( \sigma_{n+1} - \sigma_{n-1} \right) {{u}_{n}}^2
    + 2 \sum_{n=1}^{N-1} {u}_{n} {u}_{n+1}
  \right]
  + C \omega \sum_{n=1}^{N-1} (n+1) \, {u}_{n} {u}_{n+1}
  \, . \
  \label{eq:upperbound_mu_intermedio_2}
\end{equation}
Let us now observe that \eqref{eq:definizione_u} together with \eqref{eq:definizione_v} allow us to write
\begin{equation}
  \left( \sigma_{n+1} - \sigma_{n-1} \right) {u}_{n} + {u}_{n+1} + {u}_{n-1} 
  = 2 {x}_{*} {u}_{n} 
  \qquad n=1,\dots,N
  \, .
  \label{eq:successione_u}
\end{equation}
Let us multiply both sides of this last equation by ${u}_{n}$ and sum over ${n=1,\dots,N}$. 
Keeping in mind that ${{u}_{0} = 0}$ and that ${\sum_{n=1}^{N} {{u}_{n}}^{2} = 1}$, as noticed above, we obtain
\begin{equation}
  \sum_{n=1}^{N} \left( \sigma_{n+1} - \sigma_{n-1} \right) {{u}_{n}}^2
  + 2 \sum_{n=1}^{N-1} {u}_{n} {u}_{n+1}
  = 2 {x}_{*} - {u}_{N} {u}_{N+1}
  \, ,
\end{equation}
which, plugged into \eqref{eq:upperbound_mu_intermedio_2}, yields
\begin{equation}
  \mu \leq 1 
  - 2 {x}_{*} \sqrt{\dbulk_{0} \oneminus{\dbulk_{0}}} 
  + C \omega \sum_{n=1}^{N-1} (n+1) \, {u}_{n} {u}_{n+1}
  + C {u}_{N} {u}_{N+1}
  \, .
  \label{eq:upperbound_mu_intermedio_3}
\end{equation}
At this point, observing that ${\sum_{k=1}^{N} {{v}_{k}({x}_{*})}^{2} \geq {{v}_{1}({x}_{*})}^{2} = 1}$, statement (v) in Lemma~\ref{lem:successione_v} gives ${{u}_{n} \leq C {\zeta({x}_{*})}^{n}}$ for all $n$, so that we can write
\begin{equation}
  \mu \leq 1 
  - 2 {x}_{*} \sqrt{\dbulk_{0} \oneminus{\dbulk_{0}}} 
  + C \omega \sum_{n=1}^{N-1} (n+1) \, {\zeta({x}_{*})}^{2n+1}
  + C {\zeta({x}_{*})}^{2N+1}
  \, .
  \label{eq:upperbound_mu_intermedio_4}
\end{equation}
The proof is easily concluded noting that the hypothesis ${x_* > 1}$ implies ${\zeta(x_*) < 1}$, which ensures that the sum in \eqref{eq:upperbound_mu_intermedio_4} remains finite for ${N \to \infty}$.

%% file: TASEP-LK_appendix-B4.tex
\paragraph{Statement \eqref{statement1}} 

We make use of some results from the theory of difference equations, for which we refer to \cite{Agarwal00}. 
A generic sequence ${(w_n)}_{n=0}^{\infty}$ is said to be \emph{nonoscillatory} if the sequence ${({w}_{n}{w}_{n+1})}_{n=0}^{\infty}$ is eventually positive (so that ${(w_n)}_{n=0}^{\infty}$ is either eventually positive or eventually negative), and \emph{oscillatory} otherwise.
It is immediate to show that our sequence ${(v_n(x))}_{n=0}^{\infty}$ is nonoscillatory for ${x > 1}$ (see \cite{Agarwal00}, Theorem 6.5.5) and oscillatory for ${x < 1}$ (see \cite{Agarwal00}, Theorem 6.5.3). 
In order to settle the case ${x = 1}$, let us consider the sequence ${(\sigma_{n})}_{n=0}^{\infty}$ defined by \eqref{eq:definizione_sigma_ricorsiva} and, for all ${n > 0}$, let us define ${\xi_{n} \equiv \sigma_{n+1} - \sigma_{n-1}}$. 
Lemma~\ref{lem:eta} shows that ${\xi_{n} > 0}$ for all $n$ (due to the hypothesis ${\alpha < \dbulk_0}$) and ${\xi_{n} \to 0}$ for ${n \to \infty}$, with an exponential decay. Then, there must exist an integer ${m > 0}$ such that ${{(2\sum_{i=n}^{\infty} \xi_{i})}^2 \leq \xi_{n}}$ for all ${n \geq m}$. 
Defining the sequence ${(w_n)}_{n=0}^{\infty}$ as 
\begin{equation}
  w_n \equiv \cases{
	0 & if $n < m$ \\
	\textstyle \prod_{k=m}^{n} ( 1 + 2 \sum_{i=k}^{\infty} \xi_i ) 
	& if $n \geq m$
  }
  \, ,
\end{equation}
for ${n \geq m}$ we have both ${w_n > 0}$ and
\begin{equation}
  {w}_{n+1} - \left( 2 - \xi_{n} \right) {w}_{n} + {w}_{n-1} 
  \leq \left[ 
    \frac{{(2 \sum_{i=n}^{\infty} \xi_i)}^{2}}{1 + 2 \sum_{i=n}^{\infty} \xi_i} 
    - \xi_n 
  \right] {w}_{n} \leq 0
\, .
\end{equation}
At this point we can invoke Corollary 6.8.3 in \cite{Agarwal00}, which immediately proves that the sequence ${(v_n(1))}_{n=0}^{\infty}$ is nonoscillatory.

\paragraph{Statement \eqref{statement2}} 

Still with reference to \cite{Agarwal00}, we make use of the following result. 
If ${(\overline{w}_{n})}_{n=0}^{\infty}$ and ${(\underline{w}_{n})}_{n=0}^{\infty}$ are two sequences satisfying, for all ${n > 0}$, the relationships ${\overline{w}_{n+1} + \overline{w}_{n-1} = \overline{\kappa}_{n} \overline{w}_{n}}$ and ${\underline{w}_{n+1} + \underline{w}_{n-1} = \underline{\kappa}_{n} \underline{w}_{n}}$ with ${\overline{w}_{0} = \underline{w}_{0}}$ and ${\overline{w}_{1} = \underline{w}_{1}}$, and if ${\underline{w}_{n} > 0}$ and ${\overline{\kappa}_n \geq \underline{\kappa}_{n}}$ for all ${n > 0}$, then ${\overline{w}_{n} \geq \underline{w}_{n}}$ for all $n$ (see \cite{Agarwal00}, Theorem 6.8.1). 
Let us assume in particular ${\overline{w}_{0} \equiv \underline{w}_{0} \equiv 0}$ and ${\overline{w}_{1} \equiv \underline{w}_{1} \equiv 1}$. 
Setting ${\overline{\kappa}_{n} \equiv 2x - \sigma_{n+1} + \sigma_{n-1}}$ and ${\underline{\kappa}_{n} \equiv 2}$ entails ${\overline{w}_{n} = {v}_{n}(x)}$ and ${\underline{w}_{n} = n}$, so that, if $x$ is such that $2x - \sigma_{n+1} + \sigma_{n-1} \geq 2$ for all ${n > 0}$, from the above theorem we get ${v_n(x) \geq n > 0}$ for all ${n > 0}$. The required $x$ exists, because the sequence ${(\sigma_n)}_{n=0}^{\infty}$ is bounded, as it can be argued by Lemma~\ref{lem:eta}. 
Moreover, setting ${\overline{\kappa}_{n} \equiv 2y - \sigma_{n+1} + \sigma_{n-1}}$ and ${\underline{\kappa}_{n} \equiv 2x - \sigma_{n+1} + \sigma_{n-1}}$ entails ${\overline{w}_n = v_n(y)}$ and ${\underline{w}_n = v_n(x)}$, so that, if ${y \geq x}$, we get ${v_n(y) \geq v_n(x)}$ for all $n$. 
This argument shows that the set $\mathcal{X}$ of all real numbers $x$ with the property that ${v_n(x) > 0}$ for all ${n > 0}$ is an infinite interval. 
The set $\mathcal{X}$ is contained in $[1,\infty)$ because the sequence ${(v_n(x))}_{n=0}^{\infty}$ is oscillatory when ${x < 1}$, according to statement \eqref{statement1}. 
Furthermore, the set $\mathcal{X}$ is closed, because we can prove that, if a sequence ${{(x_k)}_{k=1}^{\infty} \subseteq \mathcal{X}}$ converges to $x_\infty$, then ${x_\infty \in \mathcal{X}}$. 
Indeed, we have ${v_n(x_k) > 0}$ for all ${k > 0}$ and ${n > 0}$ by hypothesis, which implies ${v_n(x_\infty) \geq 0}$ for each ${n > 0}$, once $k$ is sent to
infinity (the functions $v_n(x)$ of the variable $x$
are continuous, being polynomials). The number $x_\infty$ belongs to
$\mathcal{X}$ if the stronger condition ${v_n(x_\infty) > 0}$ is satisfied for all ${n > 0}$. 
The latter statement can be proved by contradiction as follows. 
If there were an integer ${m > 1}$ such that ${v_{m}(x_\infty) = 0}$, then, according to \eqref{eq:definizione_v}, we would find ${v_{m+1}(x_\infty) + v_{m-1}(x_\infty) = 0}$.
This would imply ${v_{m+1}(x_\infty) = v_{m-1}(x_\infty) = 0}$, because ${v_n(x_\infty) \geq 0}$ for all $n$, and therefore ${v_n(x_\infty) = 0}$ for all $n$, contradicting the fact that ${v_1(x_\infty) = 1}$.

So far we have proved that ${\mathcal{X} = [x_*,\infty)}$ with ${x_* \geq 1}$. 
In order to complete statement \eqref{statement2}, the remaining point to be proved is ${x_* < x_\circ}$, where we note that ${x_\circ > 1}$ because, according to our hypothesis \eqref{eq:intervallo_valori_q0}, we have ${\oneminus{(\beta+\Omega)} = \dbulk_0 > 1/2}$. 
As a preliminary step, we study the sequence ${(v_n(x_\circ))}_{n=0}^{\infty}$, showing that it satisfies the condition ${v_n(x_\circ) > 0}$ for all ${n > 0}$, which entails ${x_\circ \in \mathcal{X}}$ and thence the weak inequality ${x_* \leq x_\circ}$. 
Let us consider the sequence ${{(\sigma_n)}_{n=0}^{\infty}}$ defined by \eqref{eq:definizione_sigma_ricorsiva} and let us observe that for all ${n > 0}$ it is possible to write
\begin{equation}
  2x_\circ - \sigma_{n+1} + \sigma_{n-1} 
  = \frac{1}{\sigma_{n}} + \sigma_{n-1}
  \, .
  \label{eq:relazione_xo}
\end{equation}
Taking into account \eqref{eq:definizione_v}, we then obtain
\begin{equation}
  \left[ 
    {v}_{n+1}(x_\circ) - \frac{{v}_{n}(x_\circ)}{\sigma_{n}} 
  \right]
  = \sigma_{n-1} \left[ 
  {v}_{n}(x_\circ) - \frac{{v}_{n-1}(x_\circ)}{\sigma_{n-1}} 
  \right] 
  \label{eq:equazione_ricorsiva_vnxo}
\end{equation}
for all ${n > 0}$, with ${v_0(x_\circ) = 0}$ and ${v_1(x_\circ) = 1}$. 
The result that ${{v}_{n}(x_\circ) > 0}$ for all ${n > 0}$ descends from the fact that ${\sigma_n > 0}$ for all $n$ (see Lemma~\ref{lem:eta}), also noting that the terms in square brackets in \eqref{eq:equazione_ricorsiva_vnxo} are the same with shifted indices. 
As a consequence, we can prove by induction that the condition ${{v}_{n+1}(x_\circ) > {v}_{n}(x_\circ) / \sigma_{n} > 0}$, being obviously verified for ${n = 1}$, holds for all ${n > 0}$. 

Let us now also prove that there exists ${x < x_\circ}$ such that ${v_n(x) > 0}$ for all ${n > 0}$, which implies the strict inequality ${x_* < x_\circ}$. 
Let us first observe that, using Lemma~\ref{lem:eta} and \eqref{eq:intervallo_valori_gamma}, we can write
\begin{equation}
  \lim_{n \to \infty} \sigma_n 
  = \lim_{n \to \infty} 
  \frac{\ddlim_n}{\sqrt{\dbulk_0 \oneminus{\dbulk_0}}} 
  = \sqrt{\frac{\dbulk_0}{\oneminus{\dbulk_0}}}
  = \frac{1}{\sqrt{\gamma}} > 1
  \, .
\end{equation}
As a consequence, there must exist an integer ${m > 0}$ such that ${\sigma_m > 1}$. 
Let us define the function 
\begin{equation}
  c(x) \equiv 1 + 2 \left( x - x_\circ \right) 
  \left( \sigma_m - \frac{1}{\sigma_m} \right)^{-1} 
  \, .
  \label{eq:uguaglianza_x-c}
\end{equation}
Since we have previously proved that the condition ${{v}_{n+1}(x_\circ) > {v}_{n}(x_\circ) / \sigma_{n} > 0}$ holds for all ${n > 0}$, and since ${c(x_\circ) = 1}$, then by continuity there must exist ${x < x_\circ}$ (understanding $x$ close enough to $x_\circ$) such that the condition 
\begin{equation}
  {v}_{n+1}(x) > \frac{{v}_{n}(x)}{\sigma_{n} c(x)} > 0 
  \label{eq:disuguaglianza_doppia}
\end{equation}
can be satisfied up to a finite $n$, specifically for ${n=1,\dots,m}$. 
It is possible to choose $x$ in such a way that 
\begin{equation}
  x \geq x_\circ - \frac{1}{2} 
  \left( 1 - \frac{\sigma_{m\hphantom{+1}}}{\sigma_{m+1}} \right) 
  \left( \sigma_m - \frac{1}{\sigma_m} \right)
  \, ,
  \label{eq:disuguaglianza_c}
\end{equation}
because from Lemma~\ref{lem:eta}, in the hypothesis ${\alpha < \dbulk_0}$, we know that the sequence ${(\sigma_n)}_{n=0}^{\infty}$ is strictly increasing, so that ${\sigma_{m+1} > \sigma_{m} > 1}$, and therefore the right-hand side of \eqref{eq:disuguaglianza_c} is strictly smaller than $x_\circ$. 
We see that \eqref{eq:disuguaglianza_c} entails ${c(x) \geq \sigma_{m}/\sigma_{m+1}}$ and consequently, still taking into account that $\sigma_n$ increases with $n$, we can write  
\begin{equation}
  \sigma_m - \frac{1}{\sigma_m} \leq 
  \sigma_{n-1} - \frac{1}{\sigma_n c(x)} 
  \label{eq:disuguaglianza_sigma}
\end{equation}
for all ${n > m}$. 
Plugging the above inequality into \eqref{eq:uguaglianza_x-c} and observing that ${c(x) < 1}$, we get 
\begin{equation}
  2 \left(x-x_\circ\right) \geq [c(x)-1] \left[ \sigma_{n-1} - \frac{1}{\sigma_n c(x)} \right]
  \label{eq:disuguaglianza_x-c}
\end{equation}
for all ${n > m}$. 
Furthermore, taking into account also \eqref{eq:relazione_xo}, we arrive at
\begin{equation}
  2x - \sigma_{n+1} + \sigma_{n-1} 
  \geq \frac{1}{\sigma_{n} c(x)} + \sigma_{n-1} c(x)
  \label{eq:condizione_sufficiente}
\end{equation}
for all ${n > m}$. 
It is possible to show that this last inequality is a sufficient condition for \eqref{eq:disuguaglianza_doppia} to hold even for all ${n > m}$. 
Indeed, still taking into account \eqref{eq:definizione_v}, \eqref{eq:condizione_sufficiente} entails that, if for a given ${n > m}$ we have ${v_n(x) > 0}$, then  
\begin{equation}
  \left[ 
    {v}_{n+1}(x) - \frac{{v}_{n}(x)}{\sigma_{n} c(x)} 
  \right]
  \geq \sigma_{n-1} c(x) \left[ 
    {v}_{n}(x) - \frac{{v}_{n-1}(x)}{\sigma_{n-1} c(x)} 
  \right] 
  \, .
  \label{eq:disequazione_ricorsiva_vnx}
\end{equation}
We can then prove \eqref{eq:disuguaglianza_doppia} for all ${n > m}$ by induction, starting from the case ${n = m+1}$, in which we have previously proved that the condition ${{v}_{n}(x) > {v}_{n-1}(x)/[\sigma_{n-1}c(x)] > 0}$ is verified.

\paragraph{Statement \eqref{statement3}}

Let us observe that $\zeta(x)$ has been defined to be the smaller root of the characteristic equation associated with the difference equation \eqref{eq:definizione_v}, without the terms in $\sigma_n$ (that is ${\zeta} + {\zeta}^{-1} = 2x$, the other root being ${\zeta}^{-1}$). 
Then, multiplying both sides of \eqref{eq:definizione_v} by ${\zeta(x)}^{n}$ (note that ${\zeta(x) > 0}$ for all ${x \geq 1}$), by simple algebra we have
\begin{eqnarray}
  \left[
    {v}_{n+1}(x) {\zeta(x)}^{n} - {v}_{n}(x) {\zeta(x)}^{n+1}
  \right]
  - \left[
    {v}_{n}(x) {\zeta(x)}^{n-1} - {v}_{n-1}(x) {\zeta(x)}^{n}
  \right]
  \nonumber \\
  = - \left( \sigma_{n+1} - \sigma_{n-1} \right) {v}_{n}(x) {\zeta(x)}^{n} 
\end{eqnarray}
for all ${n > 0}$. 
Summing over $n$ with ${v_0(x) = 0}$ and ${v_1(x) = 1}$, dividing by ${\zeta(x)^{2n}}$ and using definition \eqref{eq:definizione_fnx}, we can derive 
\begin{eqnarray}
  \frac{{v}_{n+1}(x)}{{\zeta(x)}^{n\vphantom{1}}} 
  - \frac{{v}_{n}(x)}{{\zeta(x)}^{n-1}}
  = \frac{1 - f_n(x)}{{\zeta(x)}^{2n}}
  \label{eq:primordine}
\end{eqnarray}
for all ${n \geq 0}$. 
Note that the initial condition ${v_1(x) = 1}$ is incorporated in \eqref{eq:primordine}, since ${v_0(x) = 0}$ implies ${f_0(x) = 0}$. 
If we now sum \eqref{eq:primordine} over $n$, starting from any given $k$, we obtain 
\begin{equation}
  \frac{{v}_{n}(x)}{{\zeta(x)}^{n-1}} = \frac{{v}_{k}(x)}{{\zeta(x)}^{k-1}} 
  + \sum_{l=k}^{n-1} \frac{1 - f_l(x)}{{\zeta(x)}^{2l}}
  \label{eq:primordine_sommata}
\end{equation}
for all ${n > k}$. 

From statement \eqref{statement1} we know that, for any given ${x \geq 1}$, the sequence ${(v_n(x))}_{n=0}^{\infty}$ is non-oscillatory, meaning that it is eventually positive or negative, i.e.~there exists ${m > 0}$ such that ${v_n(x) > 0}$ or respectively ${v_n(x) < 0}$ for all ${n > m}$. 
Let us consider the positive case first.
It follows immediately from \eqref{eq:definizione_fnx} that the sequence ${(f_n(x))}_{n=0}^{\infty}$ is eventually increasing (recall that ${\sigma_{n+1}-\sigma_{n-1}}$ is always positive), that is ${f_{n}(x) > f_{n-1}(x)}$ for all ${n > m}$. 
As a consequence, using \eqref{eq:primordine_sommata} we can write
\begin{equation}
  0 < \frac{{v}_{n}(x)}{{\zeta(x)}^{n-1}} \leq 
  \frac{{v}_{k}(x)}{{\zeta(x)}^{k-1}} 
  + [1 - f_k(x)] \sum_{l=k}^{n-1} \frac{1}{{\zeta(x)}^{2l}} 
  \label{eq:primordine_disuguaglianza_1}
\end{equation}
for all ${k \geq m}$ and for all ${n > k}$, which immediately gives 
\begin{equation}
  f_k(x) < 1 + 
  \frac{{v}_{k}(x) {\zeta(x)}^{k+1}}{\sum_{l=0}^{n-k-1} {\zeta(x)}^{-2l}} 
  \label{eq:primordine_disuguaglianza_2}
\end{equation}
still for all ${k \geq m}$ and for all ${n > k}$. 
Thence, sending $n$ to infinity and observing that the denominator diverges with $n$, we obtain
\begin{equation}
  f_k(x) \leq 1 
  \label{eq:primordine_disuguaglianza_3}
\end{equation}
for all ${k \geq m}$. 
The limit \eqref{eq:definizione_fx} exists since the sequence ${(f_n(x))}_{n=0}^{\infty}$ is eventually increasing, whereas the result ${f(x) \leq 1}$ follows easily from \eqref{eq:primordine_disuguaglianza_3}. 

In the opposite case, we have ${v_n(x) < 0}$ for all ${n > m}$, and it follows immediately from \eqref{eq:definizione_fnx} that ${f_{n}(x) < f_{n-1}(x)}$ for all ${n > m}$. 
As a consequence, \eqref{eq:primordine_sommata} leads to a fully analogous argument with opposite inequalities, and thence to the result ${f(x) \geq 1}$.

\paragraph{Statement \eqref{statement4}}

From statement \eqref{statement3} we have in particular that ${f(x_*) \leq 1}$, because ${x_* \in \mathcal{X}}$, according to statement \eqref{statement2}, and thence ${v_n(x_*) > 0}$ for all ${n > 0}$. 
Here we are meant to show that ${f(x_*) = 1}$ if ${x_* > 1}$, for which it is enough to prove that ${f(x_*) \geq 1}$ within this case. 
Let us first observe that, for each integer ${k > 0}$, there exists ${\x_k \in (1,x_*)}$ such that ${v_n(x) > 0}$ for all ${x \geq \x_k}$ and all ${n = 1,\dots,k}$ (since ${x_* > 1}$ and ${v_n(x_*) > 0}$, $v_n(x)$ being continuous functions of $x$). 
It is obviously possible to choose $\x_k$ in such a way that ${\x_{k+1} \geq \x_{k}}$ for all ${k > 0}$. 
We thus have a non-decreasing sequence, being upper-bounded by $x_*$, which entails that the limit ${\x_\infty \equiv \lim_{k \to \infty} \x_k}$ exists and ${\x_\infty \leq x_*}$.  
It is in fact easy to prove by contradiction that ${\x_\infty = x_*}$ (if we had otherwise ${\x_\infty < x_*}$, this would imply ${x_* > \inf \mathcal{X}}$, contradicting statement~\eqref{statement2}). 
Let us now observe that, for any ${k > 0}$, there must exist some integer ${l \geq k}$ satisfying ${v_{l+1}(\x_k) \leq 0}$ (because ${\x_k \notin \mathcal{X}}$ as ${\x_k < x_*}$ by construction). 
Let us denote by $\ell_k$ the smallest of these integers. By construction we have ${\ell_k \geq k}$, while ${v_n(\x_k) > 0}$ for all ${n=1,\dots,\ell_k}$ and ${v_{\ell_k+1}(\x_k) \leq 0}$. 
Furthermore, using \eqref{eq:primordine} we can write, for any ${k > 0}$, 
\begin{equation}
  \frac{{v}_{\ell_k+1}(\x_k)}{{\zeta(\x_k)}^{\ell_k}} 
  - \frac{{v}_{\ell_k}(\x_k)}{{\zeta(\x_k)}^{\ell_k-1}}
  = \frac{1 - f_{\ell_k}(\x_k)}{{\zeta(\x_k)}^{2\ell_k}}
  \, .
\end{equation}
Thus, taking into account that ${v_{\ell_k+1}(\x_k) \leq 0}$ and ${v_{\ell_k}(\x_k) > 0}$, we obtain 
\begin{equation}
  f_{\ell_k}(\x_k) > 1
  \label{eq:disuguaglianza_fl}
\end{equation}
for all ${k > 0}$. 
Let us now pick any integer ${m > 0}$ and let us focus on ${k > m}$. 
Then, ${\ell_k > m}$ because ${\ell_k \geq k}$, as shown above, and by \eqref{eq:disuguaglianza_fl} and \eqref{eq:definizione_fnx} it easily follows that
\begin{equation}
  f_m(\x_k) > 1 - \sum_{n=m+1}^{\ell_k}
  \left( \sigma_{n+1} - \sigma_{n-1} \right) v_n(\x_k) {\zeta(\x_k)}^{n}
  \, .
  \label{eq:bound_fxh}
\end{equation}
The term ${v_n(\x_k) {\zeta(\x_k)}^{n}}$ can be bounded as follows. 
Since ${v_0(x) = 0}$, from \eqref{eq:primordine_sommata} we get
\begin{equation}
  \frac{{v}_{n}(x)}{{\zeta(x)}^{n-1}} = 
  \sum_{l=0}^{n-1} \frac{1 - f_l(x)}{{\zeta(x)}^{2l}}
  \label{eq:primordine_sommata_zero}
\end{equation}
for all ${n > 0}$. 
Then, choosing ${x = \x_k}$ and recalling that for all ${n=1,\dots,\ell_k}$ we have ${v_n(\x_k) > 0}$ (and thence ${f_l(\x_k) \geq 0}$ for all ${l=0,\dots,\ell_k}$), we can deduce that 
\begin{equation}
  0 < \frac{{v}_{n}(\x_k)}{{\zeta(\x_k)}^{n-1}} \leq 
  \sum_{l=0}^{n-1} \frac{1}{{\zeta(\x_k)}^{2l}} 
\end{equation}
for all ${k > 0}$ and for all ${n=1,\dots,\ell_k}$. 
Furthermore, recalling that ${\x_k > 1}$ (and thence ${\zeta(\x_k) < 1}$), multiplying by ${\zeta(\x_k)}^{2n-1}$ we easily obtain
\begin{equation}
  0 < v_n(\x_k) {\zeta(\x_k)}^{n}
  < \frac{1}{{\zeta(\x_k)}^{-1} - {\zeta(\x_k)}}
  \leq C
  \label{eq:bound_vzk}
\end{equation}
for all ${k > 0}$ and for all ${n=1,\dots,\ell_k}$. 
The last inequality can be proved observing that $\zeta(\x_k) \leq \zeta(\x_1) < 1$, which descend respectively from ${\x_k \geq \x_1 > 1}$. 
Plugging \eqref{eq:bound_vzk} into \eqref{eq:bound_fxh}, we thus obtain 
\begin{equation}
  f_m(\x_k) > 1 - C \sum_{n=m+1}^{\ell_k} \left( \sigma_{n+1} - \sigma_{n-1} \right) 
\end{equation}
for all ${m > 0}$ and ${k > m}$. 
Let us now send $k$ to infinity, and observe that ${\lim_{k \to \infty} \ell_k = \infty}$, since ${\ell_k \geq k}$. Recalling that ${\lim_{k \to \infty} \x_k = x_*}$ and that $f_m(x)$ is a continuous function of $x$, we arrive at 
\begin{equation}
  f_m(x_*) \geq 1 - C \sum_{n=m+1}^{\infty} \left( \sigma_{n+1} - \sigma_{n-1} \right) 
\end{equation}
for all ${m > 0}$.
The result ${f(x_*) \geq 1}$ is finally obtained sending $m$ to infinity.

\paragraph{Statement \eqref{statement5}}

Using \eqref{eq:primordine_sommata_zero} for ${x = x_*}$, and exploiting the fact that ${x_* > 1}$ implies ${f(x_*) = 1}$, we can write 
\begin{equation}
  \frac{v_n(x_*)}{{\zeta(x_*)}^{n-1}}
  = \sum_{l=0}^{n-1} 
  \frac{\sum_{k=l+1}^{\infty} (\sigma_{k+1} - \sigma_{k-1}) \,
  	v_k(x_*) {\zeta(x_*)}^{k}}{{\zeta(x_*)}^{2l}}
  \label{eq:vnxstar}
\end{equation}
for all ${n > 0}$. 
By an argument fully analogous to that leading to \eqref{eq:bound_vzk}, for ${x_* > 1}$ we can see that ${v_k(x_*){\zeta(x_*)}^{k} \leq C}$ for all ${k > 0}$. 
Then, also taking into account the bound ${\sigma_{k+1} - \sigma_{k-1} \leq C \gamma^k}$, which can be easily deduced from Lemma~\ref{lem:eta}, from \eqref{eq:vnxstar} together with \eqref{eq:relazione_gamma-zetaxo} we can deduce 
\begin{equation}
  \frac{v_n(x_*)}{{\zeta(x_*)}^{n-1}}
  \leq C \sum_{l=0}^{n-1} {\left[ \frac{\zeta(x_\circ)}{\zeta(x_*)} \right]}^{2l} 
\end{equation}
for all ${n > 0}$.
Now, the fact that $\zeta(x)$ is monotonically decreasing implies that the condition ${x_* < x_\circ}$, which is always satisfied by virtue of statement \eqref{statement2}, is equivalent to ${\zeta(x_*) > \zeta(x_\circ)}$. 
This is enough to prove the bound ${{{v}_{n}({x}_{*}) \leq {C}{\zeta({x}_{*})}^{n}}}$.